\documentclass[a4paper,fleqn,usenatbib,usedcolumn]{mnras}

\usepackage{mathptmx}

\usepackage[T1]{fontenc}
\usepackage{ae,aecompl}

\usepackage{graphicx}	
\usepackage{amsmath}	
\usepackage{amssymb}	
\usepackage{pdflscape}

\usepackage[dvipsnames]{xcolor}
\usepackage{xfrac}
\usepackage{algorithm2e}
\usepackage{multicol}
\usepackage{multirow}
\usepackage{bigstrut}
\usepackage[normalem]{ulem}

\usepackage[english]{babel}
\usepackage{microtype}

\pdfpageattr{/Group <</S /Transparency /I true /CS /DeviceRGB>>}

\newcommand{\textsup}[1]{\textsuperscript{#1}}

\newcommand{\forbidden}[3]{\mbox{[#1\,\textsc{#2}]{\smaller#3}}}
\newcommand{\Halpha}{\mbox{H\ensuremath{\alpha}}}
\newcommand{\Hbeta}{\mbox{H\ensuremath{\beta}}}
\newcommand{\Hgamma}{\mbox{H\ensuremath{\gamma}}}
\newcommand{\Hdelta}{\mbox{H\ensuremath{\delta}}}
\newcommand{\Hepsilon}{\mbox{H\ensuremath{\epsilon}}}
\newcommand{\angstrom}{\textup{\AA}}

\def\GA{\mathrel{\raisebox{0.13\baselineskip}{\hbox{\rlap{\hbox{\lower0.41\baselineskip\hbox{$\sim$}}}\hbox{$>$}}}}}
\def\LA{\mathrel{\raisebox{0.13\baselineskip}{\hbox{\rlap{\hbox{\lower0.41\baselineskip\hbox{$\sim$}}}\hbox{$<$}}}}}

\defcitealias{2004MNRAS.348L..59P}{PP04}
\defcitealias{2008A&A...488..463M}{M08}
\defcitealias{2013ApJS..208...10D}{D13}
\defcitealias{2017MNRAS.468.2140C}{C17}

\def\aj{AJ}
\def\araa{ARA\&A}
\def\apj{ApJ}
\def\apjl{ApJ}
\def\apjs{ApJS}
\def\aap{A\&A}
\def\mnras{MNRAS}
\def\pasp{PASP}
\def\nat{Nature}

\title[{MUSE metallicity gradients}]{First gas-phase metallicity gradients of $0.1 \LA z \LA 0.8$ galaxies with MUSE}

\author[David Carton et al.]{
David Carton,$^{1,2}$\thanks{E-mail: \href{mailto:david.carton@univ-lyon1.fr}{david.carton@univ-lyon1.fr}}
Jarle Brinchmann,$^{1,3}$
Thierry Contini,$^{4}$
Beno\^{i}t Epinat,$^{4,5}$\newauthor{}
Hayley Finley,$^{4}$
Johan Richard,$^{2}$
Vera Patr\'{i}cio,$^{2,6}$
Joop Schaye,$^{1}$\newauthor{}
Themiya Nanayakkara,$^{1}$
Peter M. Weilbacher$^{7}$ and
Lutz Wisotzki$^{7}$
\\
$^{1}$Leiden Observatory, Leiden University, PO Box 9513, 2300 RA, Leiden, The Netherlands\\
$^{2}$Univ Lyon, Univ Lyon1, Ens de Lyon, CNRS, Centre de Recherche Astrophysique de Lyon UMR5574, 69230, Saint-Genis-Laval, France\\
$^{3}$Centro de Astrofisica, Universidade do Porto, Rua das Estrelas, 4150-762, Porto, Portugal\\
$^{4}$Institut de Recherche en Astrophysique et Plan\'{e}tologie (IRAP), Universit\'{e} de Toulouse, CNRS, UPS, F-31400 Toulouse, France\\
$^{5}$Aix Marseille Univ, CNRS, CNES, LAM, Marseille, France\\
$^{6}$Dark Cosmology Centre, Niels Bohr Institute, University of Copenhagen, Juliane Maries Vej 30, 2100 Copenhagen, Denmark\\
$^{7}$Leibniz-Institut f\"{u}r Astrophysik Potsdam (AIP), An der Sternwarte 16, 14482 Potsdam, Germany
}

\date{Accepted 2018 May 16. Received 2018 May 16; in original form 2017 October 19}

\pubyear{2018}

\begin{document}
\label{firstpage}
\selectlanguage{english}
\pagerange{\pageref{firstpage}--\pageref{lastpage}}
\maketitle

\begin{abstract}
Galaxies at low-redshift typically possess negative gas-phase metallicity gradients (centres more metal-rich than their outskirts).
Whereas, it is not uncommon to observe positive metallicity gradients in higher-redshift galaxies ($z \GA 0.6$).
Bridging these epochs, we present gas-phase metallicity gradients of 84 star-forming galaxies between $0.08 < z < 0.84$.
Using the galaxies with reliably determined metallicity gradients, we measure the median metallicity gradient to be negative ($-0.039^{+0.007}_{-0.009}\, \textrm{dex}/\textrm{kpc}$).
Underlying this, however, is significant scatter: $\left(8\pm3\right)\!\%$~[7] of galaxies have significantly positive metallicity gradients, $\left(38\pm5\right)\!\%$~[32] have significantly negative gradients, $\left(31\pm5\right)\!\%$~[26] have gradients consistent with being flat.
(The remaining $\left(23\pm5\right)\!\%$~[19] have unreliable gradient estimates.)
We notice a slight trend for a more negative metallicity gradient with both increasing stellar mass and increasing star formation rate (SFR).
However, given the potential redshift and size selection effects, we do not consider these trends to be significant.
Indeed, once we normalize the SFR relative to that of the main sequence, we do not observe any trend between the metallicity gradient and the normalized SFR.
This is contrary to recent studies of galaxies at similar and higher redshifts.
We do, however, identify a novel trend between the metallicity gradient of a galaxy and its size.
Small galaxies ($r_d < 3\,\textrm{kpc}$) present a large spread in observed metallicity gradients (both negative and positive gradients).
In contrast, we find no large galaxies ($r_d > 3\,\textrm{kpc}$) with positive metallicity gradients, and overall there is less scatter in the metallicity gradient amongst the large galaxies.
These large (well-evolved) galaxies may be analogues of present-day galaxies, which also show a common negative metallicity gradient.
\end{abstract}


\begin{keywords}
galaxies: evolution -- galaxies: abundances -- galaxies: ISM
\end{keywords}



\section{Introduction}

Gas is a key ingredient for star-formation in galaxies.
Understanding how galaxies gain and lose gas is essential to explaining galaxy evolution.
We know that galaxies, both now and in the past, have insufficient gas reserves to sustain star-formation for long periods \citep{2013ApJ...768...74T}.
Consequently we know that galaxies continue to acquire gas throughout their lives.

Metals provide a chemical tag that identifies the gas that has previously been associated with star-formation.
Therefore by tracing the spatial distribution of gas-phase metallicity\footnote{Unless otherwise stated gas-phase metallicity (or simply metallicity) refers to the oxygen abundance ($12 + \log_{10}\left(\textrm{O}/\textrm{H}\right)$).} throughout a galaxy we can learn how gas is recycled and redistributed within galaxies.
Equally, we can also study how galaxies accrete and lose their gas.

In the classical inside-out picture of galaxy evolution, the inner regions of galaxies formed first from low angular momentum gas.
And with the increase of angular momentum over time, the radial scale-length of star-formation has progressed outwards in galaxies \citep{1976MNRAS.176...31L}.
Inside-out growth can explain why the centre of the Milky Way is more chemically evolved (metal-rich) than its outskirts \citep{1999A&A...350..827P}.
Moreover it can also explain why exponentially-declining radial metallicity profiles are ubiquitous in isolated and relatively massive \mbox{($\GA 10^8\,\textrm{M}_{\sun{}}$)} low-redshift galaxies \citep[e.g.][and references therein]{1992MNRAS.259..121V, 1994ApJ...420...87Z}. 

Interestingly, not only do all low-redshift star-forming galaxies present negative metallicity gradients, they also present similar slopes (when the metallicity gradient is normalized to the size of the galaxy).
For example, \citet{2014A&A...563A..49S} and \citet{2015MNRAS.448.2030H} find the $1\sigma$ scatter in the metallicity gradients between galaxies is approximately the same magnitude as the mean of the metallicity gradient. 

That said, while the radial metallicity profile of the Milky Way appears to be well-described simply by an exponentially-declining function \citep{2017MNRAS.471..987E}, there is some debate as to whether this is really true for all galaxies.
Indeed, as noted by \citet{2014A&A...563A..49S} and many others \citep[e.g.][]{2011MNRAS.415.2439R,2012ApJ...750..122B,2016A&A...585A..47M}, a significant fraction of galaxies exhibit shallow/flat metallicity profiles in their innermost and/or outermost regions.
More specifically, in a recent study of 102 massive \mbox{($\GA 10^{10}\,\textrm{M}_{\sun{}}$)} galaxies, \citet{2018A&A...609A.119S} report that only 54\% galaxies can be described by a single metallicity gradient, with the remainder exhibiting flattening either within $\LA 0.5\,r_e$ (half an effective radius) and/or beyond $\GA 1.5\,r_e$.
However, while this appears to be the prevailing picture, it should be noted that \citet{2015MNRAS.451..210C} report that the metallicity profile of some galaxies may actually \emph{steepen} in the outer (gas dominated) regions.

Nevertheless, although it appears that in some situations there are deviations from an exponential metallicity profile, it appears also that (apart from the innermost regions) most galaxies show similar exponential metallicity profiles.
At higher redshift, the assumption of exponential metallicity profiles is hard to test (due to the limited spatial resolution), but it does seem to give a reasonable description of the data.

However, it is intriguing that the common (similar slope) metallicity gradient, which is seen in low redshift galaxies, is not observed at higher redshift ($z \GA 0.6$).
Studies have found that not only was average metallicity gradient previously flatter than today, but there was also a large amount of scatter in the observed metallicity gradients \citep{2014MNRAS.443.2695S, 2016ApJ...827...74W}.

Indeed, most striking is the fact that many high-redshift galaxies have positive (inverted) metallicity gradients \citep[e.g.][]{2012A&A...539A..93Q}.
Galaxies with centres more metal poor than their outskirts are rarely, if ever, observed in the present-day Universe.
The prevailing explanation for this phenomenon is that metal-poor gas is flowing (or has flowed) into the inner regions of these galaxies.
The inflowing gas dilutes the metals, suppressing the metallicity.
The acquisition of extra gas is subsequently expected to trigger intense star formation in the galaxy.
In this regard \citet{2014MNRAS.443.2695S} identified a weak trend for elevated star-formation rates in the galaxies with flatter and inverted metallicity gradients.

There are two mechanisms that have been proposed to cause the inflow of metal-poor gas: galaxy--galaxy interactions and cold flows.
Firstly, galaxy--galaxy interactions might trigger radial flows within a galaxy's disc, transporting metal-poor gas from the outskirts to the inner regions.
At low-redshift there is observational support for this idea.
Indeed, while it is true that there is a common metallicity gradient in isolated galaxies, it has been found that non-isolated (interacting) galaxies possess significantly flatter metallicity gradients \citep{2012ApJ...753....5R}.
Furthermore, this mechanism (where mergers flatten metallicity gradients) has been demonstrated in numerical simulations \citep{2010ApJ...710L.156R, 2012ApJ...746..108T}.
It appears, however, that galaxy--galaxy interactions are merely capable of flattening the metallicity gradient of galaxies, but not inverting it.

On the other hand, cold flows (the other mechanism proposed for producing inverted gradients), may be more successful \citep{2010Natur.467..811C}.
These flows are cold streams of gas which can penetrate through a galaxy's hot halo to reach the galaxy itself \citep{2005MNRAS.363....2K, 2006MNRAS.368....2D}.
However, it is not clear that this metal poor gas would flow directly to the centre of the galaxy, as would be needed to invert the metallicity gradient.
Indeed it has been suggested that these streams may build an extended gas disc \citep{2011ApJ...738...39S, 2015MNRAS.449.2087D}, which itself could, in turn, contract to form a compact star-forming clump at the centre of the galaxy \citep{2014MNRAS.438.1870D, 2015MNRAS.450.2327Z}.
There is some observational support for this, with the tentative identification of such cold-flow discs in a couple of high redshift galaxies \citep{2013Sci...341...50B, 2016ApJ...820..121B}.

Simulations predict that these hypothesized cold flows would dominate the gas supply of galaxies in the early Universe ($z \GA 1.8$) but we should, however, expect them to become increasingly rarer at late times \citep[e.g.][]{2011MNRAS.415.2782V, 2014MNRAS.442..732W}.
So, while cold flows may explain why galaxies observed at $z \approx 3.4$ present inverted metallicity gradients \citep{2014A&A...563A..58T}, it is harder to invoke cold flows as an explanation for inverted gradients observed at $z \approx 1$.

To summarize briefly, there is a disparity between metallicity gradients in the high-redshift and low-redshift Universe.
While not necessarily contradictory, the high-redshift results point to stochastic processes dominating galaxy evolution, whilst the low-redshift results suggest a secular evolution of galaxies.
There are few, if any, observations of metallicity gradients in galaxies between $0.1 \LA z \LA 0.6$.
Clearly, bridging this gap is a necessary step towards understanding the disparity between the high- and low-redshift results.
Here, with the Multi Unit Spectroscopic Explorer \citep[MUSE;][]{2010SPIE.7735E..08B, 2017A&A...608A...1B}), we will provide for the first time a large sample of metallicity gradients in intermediate-redshift galaxies ($0.08 < z < 0.84$).

Analysing these observations presents several challenges.
The first challenge is to correct for the effects of seeing on our data.
As demonstrated by both \citet{2013ApJ...767..106Y} and \citet{2014A&A...561A.129M}, failing to correct for seeing effects will produce systemically flatter metallicity gradients.
But the challenge posed by seeing is not unique to our work, and accordingly other recent studies have used simulated observations to apply an \emph{a posteriori correction} to infer the true metallicity gradient.
In \citet[][hereafter C17]{2017MNRAS.468.2140C}, however, we presented an alternative \emph{forward-modelling} approach.
This method is better able to quantify the degeneracies that arise from seeing-corrupted data, and therefore yields formal estimates for uncertainty in the recovered metallicity gradient.

A second challenge we face is that we derive metallicities from a combination of nebular emission-lines.
Depending on a galaxy's redshift, different emission-lines fall within the wavelength range of a spectrograph.
It is a well-documented issue that different metallicity calibrations (especially when using different emission-lines) produce different results \citep[e.g.][]{2008ApJ...681.1183K}.
With a forward-modelling approach we can overcome these limitations and thereby self-consistently infer metallicity gradients independently of redshift.

With our observations of intermediate-redshift galaxies we will attempt to reconcile the high- and low-redshift pictures of galaxy evolution, particularly with respect to the gas supply in these systems.
We structure the paper as follows.
In Section~\ref{sec:data} we describe our observations and outline our galaxy sample selection.
We detail our methodology in Section~\ref{sec:analysis}, where we also include a sensitivity analysis for our model.
Section~\ref{sec:results} is dedicated to presenting the results on the derived metallicity gradients.
In Section~\ref{sec:discussion} we provide a discussion of these results.
Finally we conclude our findings in Section~\ref{sec:conclusions}.

Throughout the paper we assume a \mbox{$\Lambda\textrm{CDM}$} cosmology with $H_0 = 70\ \textrm{km}\,\textrm{s}^{-1}\,\textrm{Mpc}^{-1}$, $\Omega_{\mathrm{m}} = 0.3$ and $\Omega_{\Lambda} = 0.7$.

\section{Data}\label{sec:data}

We wish to spatially resolve metallicity gradients in distant ($0.1 \LA z \LA 0.8$) galaxies.
Using integral-field spectroscopy (IFS) we can map the nebular emission from star forming regions in these galaxies, and therefore measure radial metallicity variations.

Here we will use observations taken with the MUSE instrument situated at UT4 of the Very Large Telescope (VLT).
We will construct our galaxy sample by combining data from both Guaranteed Time Observations (GTO) programmes and commissioning activities.
However, because of the differing observing strategies employed in these observing campaigns, our data sample is rather inhomogeneous.
Galaxies were observed with a variety of integration times (between 1 -- 31\,h) and in a variety of seeing conditions.
We will describe these datasets fully in Section~\ref{sec:field_desc}.

\subsection{MUSE Observations}

\subsubsection{Instrument Description}

MUSE is an integral-field spectrograph that employs an image slicing technique at optical wavelengths.
In normal (non-extended) wide-field mode, MUSE provides spectra over a continuous $1' \times 1'$ Field of View (FoV) with continuous spectral coverage (4750\AA{} -- 9300\AA{}).
These spectra have a wavelength resolution of $\approx2.5\textrm{\AA{}}$ full-width half-maximum (FWHM), although this is not entirely constant with wavelength \citep[see][]{2017A&A...608A...1B}.
The spatial sampling of the data is $0.2'' \times 0.2''$, but in actuality the spatial resolution of our data limited by the seeing.

\subsubsection{Field Description}\label{sec:field_desc}

Given MUSE's large contiguous field, we do not target individual galaxies, rather we target collections of galaxies with limited pre-selection.
While each field was chosen to optimize the scientific objectives of the different observing programmes, in general the galaxy selection is essentially blind.
There is one exception where one field (CGR30) targets a galaxy group at $z\approx0.7$.
But even then there will be foreground and background galaxies in this field that are blindly selected.
In total there are 35 MUSE pointings, covering $\sim 35\,\textrm{arcmin}^2$ on the sky.

While the parent galaxy sample selection is essentially blind, there are no straightforward criteria for selecting the galaxies for which we can measure metallicity gradients.
That said, we would expect that we can measure metallicity gradients in the largest and brightest galaxies at a given redshift.
We therefore do not impose a priori selection criteria, and analyse all MUSE detected galaxies that have known redshifts $z<0.9$, rejecting those with insufficient signal-to-noise (S/N).
We describe this S/N cut in Section~\ref{sec:selection_criteria}.
We present a post hoc description of the final sample in Section~\ref{sec:sample_selection}.

We will now outline the data used in our analysis as follows (a summary is displayed in Table~\ref{tab:field_descr}):
\begin{description}
\item[\textbf{Hubble Deep Field South (HDFS):}] As one of the commissioning activities of MUSE, a single deep ($26.5\,\textrm{h} = 53 \times 1800\,\textrm{s}$) field in the HDFS was acquired.
The average seeing conditions were good ($\textrm{FWHM}=0.66''$ at 7000\AA{}).
\citet{2015A&A...575A..75B} present a full description of the data.
Here we use a slightly improved data reduction to the one presented therein.
This new reduction (version 1.24) includes improvements to the sky subtraction and slice normalization (quasi flat-fielding), see \citet{2017A&A...608A...1B}.

\item[\textbf{Hubble Ultra Deep Field (UDF):}] The MUSE-Deep GTO survey has observed a 9 field mosaic that covers the UDF.
This $3' \times 3'$ field has been observed to a depth of \mbox{$\approx 10\,\textrm{h}$} (in exposures of $1500\,\textrm{s}$ each).
In addition, there is also an extra-deep $1' \times 1'$ portion of the mosaic that reaches \mbox{$\approx 31\,\textrm{h}$}.
During the observations the average seeing conditions were good, resulting in a final combined PSF with $\textrm{FWHM}=0.61''\,\textrm{--}\,0.67''$ at 7000\AA{}.
This data will appear in \citet{2017A&A...608A...1B} and the full catalogue, including redshifts and line fluxes etc., in \citet{2017A&A...608A...2I}.
Here we are using a slightly older data reduction (version 0.31) than that presented in the aforementioned papers.
The subsequent improvements in the reduction primary focus on removing low-level systematics, however, since we concern ourselves here with only the brightest galaxies these improvements are by and large irrelevant for our science.

\item[\textbf{Chandra Deep Field South (CDFS):}] The MUSE-Wide GTO program is surveying a portion of the Chandra Deep Field South (amongst other fields).
In the end this will produce a 60 tile mosaic of the CDFS at 1\,h depth (using exposures of $4 \times 900\,\textrm{s}$).
Here we will use only the first 24 fields that have been observed.
These observations were performed in moderate and poor seeing conditions, resulting in a $\textrm{FWHM}=0.7''\,\textrm{--}\,1.1''$ at 7000\AA{})
This dataset (version 1.0) is described by \citet{2017A&A...606A..12H} and in a forthcoming data release (Urrutia et al., in prep.).

\item[\textbf{COSMOS Group 30 (CGR30):}] A third GTO program is surveying galaxies in group environments.
In our analysis here we will use observations of one of these galaxy groups (namely Group~30 as identified in the zCOSMOS 20k Group Catalogue \citet{2012ApJ...753..121K}).
The deepest portion of the field covers slightly less than the full $1' \times 1'$ FoV and reaches a 9.75\,h depth ($39 \times 900\,\textrm{s}$).
However, one galaxy in our final sample (COSMOS2015 ID: 506103) lies in a shallow ``snapshot'' region and therefore was only observed to 1\,h depth.
The average seeing conditions were good ($\textrm{FWHM}=0.60''$ at 7000\AA{}).
This field (version 1.0) was presented in \citet{2018A&A...609A..40E}.

\end{description}

\begin{table}
\caption{Summary of galaxy observations.
The final sample of galaxies was obtained from various targeted fields (with differing exposure depths and seeing conditions).
We list the number of galaxies obtained from each field.
}
\label{tab:field_descr}
\begin{tabular}{cccc}
\hline
Field & Depth & PSF FWHM & \# of galaxies \\
 & $\left[\textrm{h}\right]$ & $\left[\textrm{arcsec}\right]$ & in final sample\\
\hline
HDFS & 26.5 & 0.66 & 12 \\
UDF-Medium & $\approx10$ & 0.61 -- 0.67 & 28 \\
UDF-Deep & $\approx31$ & 0.65 & 9 \\
CDFS & 1 & 0.7 -- 1.1 & 30 \\
CGR30 & 9.75 & 0.60 & 4 \\
CGR30-Snapshot & 1 & 0.60 & 1 \\
\hline
\end{tabular}
\end{table}

\subsubsection{Data Reduction}

Above we described fields from four different observing programmes and as is to be expected, there are differences in the specifics for each of the data reductions.
However, all reductions use the standard data reduction pipeline (Weilbacher, in prep.)\footnote{A short description of the pipeline can be found in \citet{2012SPIE.8451E..0BW}.} to produce calibrated datacubes.
In all fields sky subtraction is performed using the \textsc{Zurich Atmospheric Purge} \citep[\textsc{ZAP};][]{2016MNRAS.458.3210S} software, which employs a principal component analysis technique developed specifically for  MUSE data.

The largest difference between the reductions are the implementations (or lack thereof) of slice normalization.
Slice\footnote{Slice refers to the optical image slicers within MUSE, not the wavelength layers (channels).} normalization improves the uniformity (flatness) of the field and is primarily required because the flat-field calibrations are not taken at the exact same time as the science exposures.
Small changes in the instrument alignment due to thermal variations can alter the throughput to the slits.
Slice normalization is essentially a secondary flat-fielding, that self-calibrates using the individual science exposures.
Because multiple exposures are combined, these semi-random slice systematics contribute to the effective noise in the final datacube.
As a result, the application of slice normalization is very important for faint galaxies, but will have little impact on the bright galaxies that we study here.
Thus the fact that the various data reductions implement the normalization differently will not affect our results.

The final datacubes are constructed with equal sized voxels\footnote{Volumetric pixels.} ($0.2'' \times 0.2'' \times 1.25\angstrom{}$), mirroring the native pixel size at the charge-coupled device (CCD) level.
With this voxel size, the typical seeing-limited point-spread function (PSF) is well sampled, whilst the 2.5\AA{} line-spread function (LSF) is only just critically sampled.

\subsubsection{PSF Determination}

A critical part of our analysis is to forward model the seeing effects on our data.
It is therefore necessary to measure the final PSF directly from our datacubes.
MUSE commissioning activities established that the PSF is relatively spatially invariant across the FoV; a PSF model fit to a bright star somewhere within the FoV can be applied across the whole field.
Unfortunately, not all fields contain such bright stars, and therefore we use a variety of PSF determination techniques in the different fields:
\begin{description}
\item[\textbf{HDFS:}] This field contains a bright star to which \citet{2015A&A...575A..75B} fit a Moffat function.
The FWHM of the Moffat profile is allowed to vary as a function of wavelength, but the Moffat $\beta$ parameter is not.
We describe the FWHM as a piecewise linear function with three knots \{4750\AA{}, 7000\AA{}, 9300\AA{}\}.

\item[\textbf{UDF:}] Most of the MUSE UDF pointings do not contain any bright stars, in which case the PSF must be inferred from non-point source objects (i.e. galaxies).
Hubble Space Telescope (HST) images are convolved with a Moffat function and fit\footnote{The fit is performed in the Fourier space.} to MUSE pseudo-broadband images, see \citet{2017A&A...608A...1B}.
We obtain a best-fit Moffat profile as a linear function of wavelength.
The Moffat $\beta$ parameter is assumed to be constant.
The accuracy of this method has been verified by comparing the results in those fields that do contain bright stars.

\item[\textbf{CDFS:}] As with the UDF, many of the CDFS fields are also devoid of bright stars.
For these fields, \citet{2017A&A...606A..12H} determined the PSF using bright, compact galaxies.
Choosing only galaxies with minimal substructure, they parametrized the HST images of these galaxies as 2D elliptical Gaussian functions.
By convolving the galaxy models with a circular Gaussian PSF, they then derived the PSF that best matches a series of MUSE pseudo-broadband images (constructed at various wavelengths).
From this they were able to construct a PSF model for each field, where the Gaussian FWHM varied linearly as a function of wavelength.
There were, however, some fields that also contained a bright star, enabling the PSF to be additionally derived by a direct fit to this point source.
In these fields, and where the FWHM was a closer match to the telescope autoguider measurement, they adopted the stellar PSF fit in place of that fit to the galaxies.

\item[\textbf{CGR30:}] This field contains four relatively faint stars.
We perform a simultaneous fit to all stars using a Moffat PSF.
The FWHM is assumed to have a 3\textsup{rd} order polynomial dependence as a function of wavelength, whilst for the Moffat $\beta$ parameter we allow only a linear dependence with wavelength.

\end{description}

Of the four targeted areas, determining a reliable PSF model for CDFS fields was the most challenging.
Thus, to check the validity of our procedure, we applied the same method as used for the UDF to the CDFS data; we found that both methods yielded similar results.

To summarize, we model the PSF as an axisymmetric function (either a Moffat or Gaussian function).
The FWHM is free to vary as function of wavelength, and is typically found to be $\sim 25\%$ larger in the blue than in the red.
These wavelength-dependent PSF models are directly used in the forward-modelling of our observations.

\subsection{Derived global properties}

As part of our analysis, we study metallicity gradients as a function of global galaxy properties, e.g. stellar mass, star formation rate, and disc size.
We now outline how these quantities are derived.

\subsubsection{Stellar Mass}\label{sec:stellar_mass}

There exists extensive broadband photometry for all of the fields that we study here.
Stellar masses are estimated though stellar population synthesis (SPS) modelling.
This yields a self-consistent mass estimate, despite the differing availability of filters in different fields.
We use \textsc{magphys} \citep{2008MNRAS.388.1595D} to fit the photometry of each galaxy, adopting the \citet{2003MNRAS.344.1000B} SPS models with a Chabrier initial mass function \citep[IMF;][]{2003PASP..115..763C}, we also fix the redshift of the galaxy to that derived from the MUSE spectra.
These models, however, do not include any potential contribution from nebular emission lines.
To summarize, \textsc{magphys} describes galaxies with an exponentially declining star-formation history, with random bursts superimposed.
The stellar light is attenuated with a \citet{2000ApJ...539..718C} dust model, and the absorbed energy is self-consistently re-radiated in the infrared.

The photometry used in each of the four fields is derived from various sources, all of which are approximations of the total magnitude:

\begin{description}

\item[\textbf{HDFS:}] For this field we use the four-band HST photometry \{F300W, F450W, F606W, F814W\} from \citet{2000AJ....120.2747C}.

\item[\textbf{UDF:}] Extensive deep multi-band HST photometry is provied by \citet{2015AJ....150...31R} in \{F225W, F336W, F435W, F606W, F775W, F850LP, F105W, F125W, F140W, F160W\}. Where possible, we use all filters.

\item[\textbf{CDFS:}] Here we use the photometric catalogue of \citet{2013ApJS..207...24G} using exclusively the HST photometry \{F606W, F775W, F814W, F850LP, F105W, F125W, F160W\}, using all where available.

\item[\textbf{CGR30:}] In this field we adopt the photometric catalogue of \citet{2007ApJS..172...99C} using \{Subaru $B_j\,V_j\,g^+\,r^+\,i^+\,z^+\,\textrm{NB816}$, SDSS $u\,g\,r\,i\,z$, CFHT $u^{\ast}\,i^{\ast}$, HST F814W, CTIO/KPNO $K_s$\}.
\end{description}

To test the sensitivity of our results to the choice of stellar mass estimates, we also derive them using \textsc{fast} \citep{2009ApJ...700..221K}, assuming  both an exponentially declining star-formation history and adopting a \citet{2000ApJ...533..682C} dust law.
And although we identified some small differences between the \textsc{FAST} and \textsc{MAGPHYS} stellar mass estimates ($\LA 0.2\,\textrm{dex}$), we find that neither our results nor conclusions are impacted by our choice of stellar mass estimate.

\subsubsection{Star Formation Rate (SFR)}

We derive global star formation rates directly from the MUSE data, taking the spectrum integrated across the whole galaxy.
On this we perform a full spectral-fitting using \textsc{platefit} \citep{2004ApJ...613..898T,2004MNRAS.351.1151B}.
We will describe the spectral fitting in Section~\ref{sec:spectral_fitting}.
Here it suffices to say that we obtain the \Halpha{}, \Hbeta{} and \Hgamma{} emission-line fluxes, accounting for the underlying stellar absorption.

For low-redshift  galaxies ($z \LA 0.4$) we use \Halpha{} and \Hbeta{} to compute the SFR.
At higher redshifts \Halpha{} is redshifted beyond the MUSE wavelength range, so we compute the SFR in these galaxies using \Hbeta{} and \Hgamma{} instead.

To correct for dust, we adopt the \citet{2000ApJ...539..718C} birth-cloud absorption curve which attenuates the luminosity, $L(\lambda)$ as follows
\begin{equation}
L_\mathrm{ext}(\lambda) = L(\lambda) e^{-\tau\left(\lambda\right)},
\label{eq:cf00_dust_a}
\end{equation}
with
\begin{equation}
\tau(\lambda) = \tau_V \left(\frac{\lambda}{5500\,\angstrom}\right)^{-1.3},
\label{eq:cf00_dust_b}
\end{equation}
where $\tau_V$ is the V-band optical depth.
Depending on the redshift we use either the observed $\Halpha{}/\Hbeta{}$ or $\Hgamma{}/\Hbeta{}$ ratios to calculate $\tau_V$.
For this we assume intrinsic Case B Balmer recombination ratios of $j_{\textrm{H}\alpha}/j_{\textrm{H}\beta} = 2.86$ and $j_{\textrm{H}\gamma}/j_{\textrm{H}\beta} = 0.468$.
These values are appropriate for \ion{H}{ii} regions of temperatures, $T_e=10,{}000\,\textrm{K}$, and electron densities, $n_e = 100\,\textrm{cm}^{-3}$ \citep{2003adu..book.....D}.

Finally, we convert the inferred dust-corrected \Halpha{} luminosities to SFRs using a scaling relation between \Halpha{} and SFR
\begin{equation}\label{eq:halpha_sfr_law}
\log_{10} \left( \frac{\textrm{SFR}} {\textrm{M}_{\sun}\,\mathrm{yr}^{-1}} \right) = \log_{10} \left( \frac{L(\textrm{H}\alpha)}{\mathrm{erg}\,\mathrm{s}^{-1}} \right) - 41.27,
\end{equation}
\citep{2011ApJ...737...67M, 2011ApJ...741..124H, 2012ARA&A..50..531K}.
This assumes a Kroupa IMF \citep{2001MNRAS.322..231K}, and is therefore effectively consistent with the Chabrier IMF we adopted when obtaining stellar masses.

For \mbox{$z \GA 0.4$} galaxies (where \Halpha{} is not available), we assume $L(\textrm{H}\alpha) = 2.86\ L(\textrm{H}\beta)$ (i.e. the Case B ratio).

\subsubsection{Galaxy Morphology}\label{sec:morphology}

In \citetalias{2017MNRAS.468.2140C} we presented a method for modelling the metallicity gradients in our galaxies.
As inputs, this method requires four basic morphological parameters: the galaxy centre (Right Ascension, RA, and Declination, Dec.), the inclination of the galaxy (inc.) and the position angle of the major axis on the sky (PA).
For our discussion we also need galaxy size, which we shall express as the  exponential disc scale-length, $r_d$.

All the fields we present here are well studied and have existing morphological catalogues.

\begin{description}

\item[\textbf{HDFS:}] For this field \citet{2016A&A...591A..49C} provide a detailed morphological analysis of the resolved galaxies.
They perform a bulge-disc decomposition on the HST F814W imaging, which yields all the necessary morphology information.

\item[\textbf{UDF\,+\,CDFS:}] For both these fields \citet{2012ApJS..203...24V} provide a catalogue of single S\'{e}rsic fits.
While this catalogue provides most of the relevant information, it does not provide estimates for the galaxy inclinations.
The catalogue only provides the axis ratios of the galaxies.

Since axis ratio is primarily a function of inclination, we can convert axis ratios into inclinations.
However, axis ratios have a secondary dependence on the morphological type of the galaxy.
Since the S\'{e}rsic index is a proxy for morphological type, we can reduce this secondary dependence by partitioning the catalogue into quintile bins of the S\'{e}rsic index.
Within each of the these bins we order the galaxies by decreasing axis-ratio and use the rank order of the galaxy within the bin to estimate its inclination.
For this we assume that we view galaxies from a random orientation on a 3D sphere, and therefore the probability of observing a galaxy with an given inclination, $i$, is $P\left(i\right) \propto \sin\left(i\right)$.
In other words we are more likely to observe a galaxy edge-on that we are to see it exactly face-on.

To estimate the galaxy sizes we use the half-light radii, $r_e$, reported in the catalogue.
To convert these to disc scale-lengths we simply assume that the galaxy profile is a bulgeless exponential disc (i.e. $r_d \approx 0.596\,r_e$).
Note that as a result of this assumption, we will likely underestimate the true disc size.
In fact, because the most massive galaxies tend to have more prominent bulges, this might create an unwelcome trend such that the sizes of the most massive galaxies are more significantly underestimated.
Fortunately, to a great extent the correlation between stellar mass and bulge fraction is driven by the large number of passive (and bulgy) galaxies at high masses \citep{2014MNRAS.441..599B, 2015ApJ...809...95H}.
And since we will be selecting galaxies with bright nebular emission, we preferentially select against such bulgy galaxies.
As a result, we do not consider a potential mass trend to be of great concern, neither do we anticipate extremely bulge dominated systems.

The catalogue provides the morphology derived from three HST bands (F105W, F125W and F160W).
For any given galaxy we use the morphology of the band with the highest S/N.

Finally, note that while there are other \emph{optical} morphological catalogues available, we choose to use the \citet{2012ApJS..203...24V} catalogue because the near infrared data is less influenced by recent star formation activity and therefore should provide a better description for the morphology of the overall stellar population.

\item[\textbf{CGR30:}] For this field we use the morphological assessment provided by the COSMOS 2005 Morphology Catalogue, which uses \textsc{morpheus} \citep{2007ApJ...669..184A} to measure the morphological parameters.
As above, this catalogue also only reports a galaxy's axis ratio, not inclination.
We again apply the rank ordering method to convert to axis ratio into an inclination.
To avoid mixing morphological types, we use the concentration index\footnote{The ratio between the radii that contain 50\% and 90\% of a galaxies light.} as a proxy for morphological type.
We divide the catalogue into decile bins of the concentration index, and perform the rank ordering within each.
This catalogue provides the half-light radii of the galaxies, which we convert to exponential disc scale-lengths as above.

\end{description}

The different galaxy morphology catalogues are measured in different photometric bands.
This will systematically affect the measured galaxy sizes; sizes of late-type galaxies measured at redder rest-frame wavelengths will systematically appear smaller.
We correct for this using the parametrization of \citet{2014ApJ...788...28V} (their equations~1~\&~2).
This correction depends only on the galaxy's redshift and stellar mass.
In this work all galaxy sizes are quoted as if they were measured at a rest-frame wavelength of 5000\AA{}.

\subsection{Sample Description}\label{sec:sample_selection} 

As mentioned previously, we do not make an a priori selection for our sample.
Nevertheless, there are many galaxies for which we cannot meaningfully constrain the metallicity gradient.
Naturally we would expect that we can only measure metallicity gradients in the largest and brightest galaxies.
However, it is non-trivial to map this into a clean cut on global properties (e.g. stellar mass, size and SFR).
Therefore we build our final sample based upon data-driven criteria (i.e. S/N).

\subsubsection{Selection Criteria}\label{sec:selection_criteria}

We extract spatially resolved emission-line fluxes from our parent sample of 590 MUSE detected galaxies (and with MUSE redshift measurements $z < 0.85$).
The procedure for this extraction is described in Section~\ref{sec:flux_extraction}.
In many galaxies we fail to detect any emission-line component.

To meaningfully constrain metallicity (and distinguish its effects from dust) we need two strong forbidden lines and two Balmer lines to be detected at $\textrm{S/N} \ge 5$.
Exactly which emission lines are chosen depends on the galaxy.
Using the globally integrated spectrum we choose the two forbidden lines with the highest S/N, and the two Balmer lines with the highest S/N.
A typical choice for a low-redshift galaxy \mbox{($z \LA 0.4$)} is \{\Hbeta{}, \forbidden{O}{iii}{5007}, \Halpha{}, \forbidden{S}{ii}{6717,6731}\}.
And a typical choice at higher redshift is \{\forbidden{O}{ii}{3726,3729}, \Hgamma{}, \Hbeta{}, \forbidden{O}{iii}{5007}\}.

Since we need to constrain the metallicity gradient of the galaxy, not just its metallicity, the line emission must be detected in multiple spatial bins.
Explicitly we require that the four chosen emission-lines are all detected at $\textrm{S/N} \ge 5$ in at least three spatial bins.
(The spatial binning scheme is described in Section~\ref{sec:spatial_binning}.)

Overall these criteria amount to a minimum S/N cut.
How this selection maps into galaxy properties depends on the field (because our observations have different depths and seeing conditions).

In addition to this main selection cut, we apply three further criteria:

\begin{itemize}

\item In rare occasions the emission we detect may not be associated with the target of interest, i.e. the data is contaminated by a brighter neighbouring galaxy, at the same redshift.
These cases can be identified through visual inspection.
We manually excluded galaxies where significant contamination is apparent.

\item When modelling the metallicity gradients in our galaxies we assume the galaxies to be infinitesimally thin discs.
This approximation is acceptable for face-on galaxies, but, however, becomes increasingly questionable for more inclined systems.
We therefore exclude galaxies with an estimated inclination $\textrm{inc.} > 70\degr{}$.

\item Galaxies with bright active galactic nuclei (AGN) will produce bright line-emission.
Such emission would alter the observed emission-line ratios, and thus alter the inferred metallicities.
In the following section we explain how we exclude AGN.

\end{itemize}

\subsubsection{AGN Exclusion}\label{sec:agn_exclusion}

\begin{figure*}
\includegraphics[width=\linewidth]{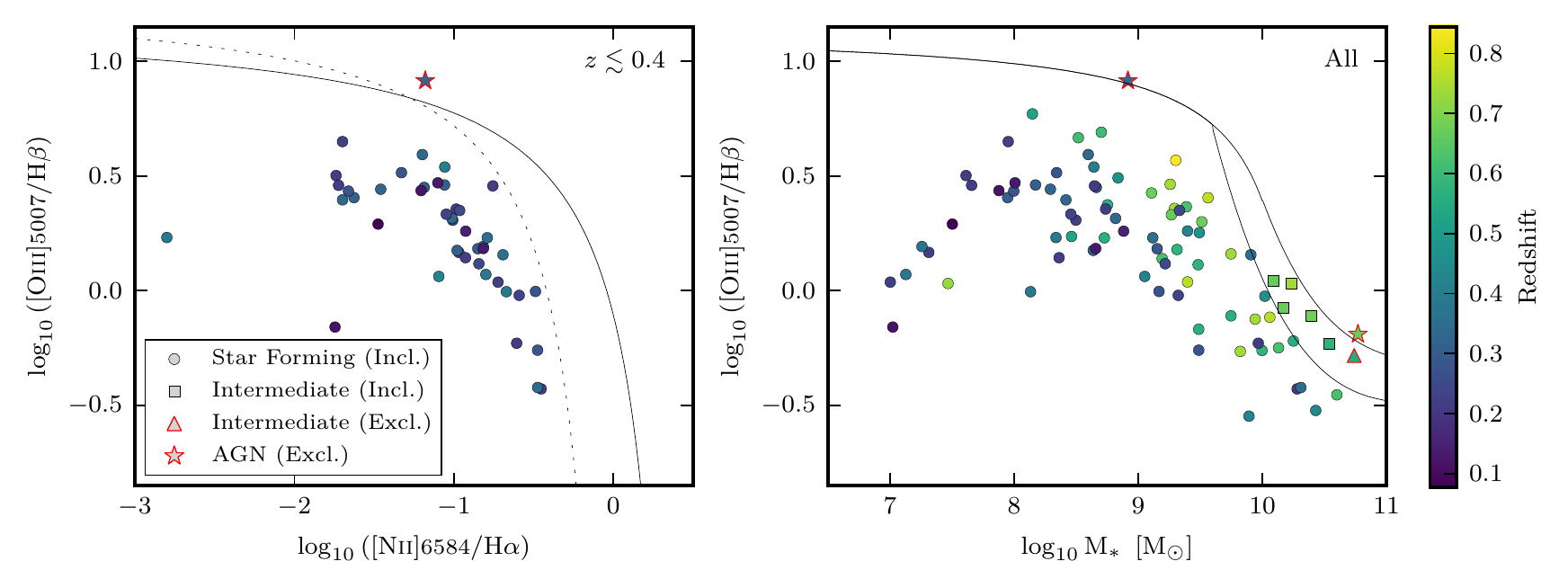}
\caption{Diagnostic plots for AGN classification.
Our final sample of star forming galaxies and included intermediate galaxies are plotted as circles and squares, respectively.
Excluded intermediate and AGN/LINER galaxies are indicated by triangles and stars, respectively.
Data points are colour coded by redshift.
(Left) The $\textrm{\forbidden{O}{iii}{}} / \textrm{H}\beta$ versus $\textrm{\forbidden{N}{ii}{}} / \textrm{H}\alpha$ diagnostic (a BPT \citep{1981PASP...93....5B} diagram) for the low-redshift ($z\LA0.4$) portion of our sample.
As a solid line we indicate the \citet{2001ApJ...556..121K} theoretical maximum limit for star formation alone.
With a dashed line we show the \citet{2003MNRAS.346.1055K} curve, the empirical division between emission from star formation and AGN/LINER emission.
(Right) The Mass-Excitation (MEx) diagram for whole our sample, including also the \citet{2014ApJ...788...88J} demarcations as solid black lines.
Note that the same galaxies that appear in the left panel (the BPT diagram) also appear in the right panel (the MEx diagram).}
\label{fig:BPT_MEx}
\end{figure*}

It is difficult to determine the metallicity of galaxies when they are contaminated by emission from AGN or low-ionization nuclear emission-line regions (LINERs).
We could treat such galaxies in one of three ways.
Firstly, one could attempt to model the flux contribution from a compact central source.
Secondly, we could mask out the central regions of a galaxy, and derive the metallicity gradient from the outer regions of the galaxies.
A third approach, and the one we adopt here, is to simply discard galaxies from our sample if they appear to have a significant AGN/LINER component.
We classify our galaxies using standard emission-line ratios classifications, which we apply to a galaxy's globally integrated spectrum.

At low redshift ($z \LA 0.4$) we use both the $\textrm{\forbidden{N}{ii}{}} / \textrm{H}\alpha$ and $\textrm{\forbidden{O}{iii}{}} / \textrm{H}\beta$ ratios to classify galaxies.
We follow the classification scheme of \citet{2004MNRAS.351.1151B} to divide galaxies into three categories: pure star forming, those with significant AGN, and those that fall in between (i.e. intermediate).
We exclude galaxies classified as AGN galaxies, but we do not automatically exclude the intermediate case galaxies.
These intermediate galaxies are inspected manually and we exclude those that clearly possess broad emission-line velocity components, indicative of AGN galaxies.

At $z \GA 0.4$ the $\textrm{\forbidden{N}{ii}{}} / \textrm{H}\alpha$ is redshifted out of MUSE's wavelength range.
We therefore adopt the Mass-Excitation (MEx) diagnostic \citep{2014ApJ...788...88J} to classify the galaxies into the same three classifications (star forming, intermediate, AGN).
As before, intermediate cases are manually inspected.

In Fig.~\ref{fig:BPT_MEx} we show where our galaxies lie with respect to the two diagnostics.
Out of 87 galaxies we exclude one galaxy below $z \approx 0.4$ and two galaxies above.

\subsubsection{Contamination from evolved stars}

In some galaxies hot low-mass evolved stars (HOLMES) can contribute a significant number of ionizing photons, giving rise to LINER-like line emission \citep{2008MNRAS.391L..29S}.
And, therefore, these stars present another possible source contamination which could affect our metallicity measurements.

In the following section (\ref{sec:sample_properties}) we shall demonstrate that our galaxy selection criteria ultimately result in a preferential selection of galaxies with high specific star formation rates.
As a result, it is a priori unlikely that, on the whole, the evolved population would dominate over the young as the main source of line emission.
However, if the star formation is not uniform, it is entirely plausible that \emph{local} regions within the galaxies could be dominated by evolved stars.

Fortunately, \citet{2011MNRAS.413.1687C} have proposed a simple cut based on the equivalent width (EW) of \Halpha{} that allows one to classify galaxies with emission dominated by the evolved stellar population; galaxies with $\textrm{EW}\left(\textrm{H}\alpha\right) > -3\,\angstrom{}$ are to be considered dominated by evolved stars\footnote{We adopt the sign convention that expresses emission as negative equivalent widths.}.
Unfortunately, we are not able to observe \Halpha{} for the highest redshift galaxies in our sample ($z \GA 0.4$).
However, all else being equal, and assuming  a case B ratio of $L(\textrm{H}\alpha) \sim 3\ L(\textrm{H}\beta)$, we can define a new cut $\textrm{EW}\left(\textrm{H}\beta\right) > -1\,\angstrom{}$.

In our analysis (Section~\ref{sec:flux_extraction}) we measure the line emission in our galaxies divided into 2445 spatial bins.
Of these, only 54 ($\sim 2\%$) of the bins have $\textrm{EW}\left(\textrm{H}\beta\right) > -1\,\angstrom{}$.
Note that there is also a large uncertainly associated with these measurements (\mbox{$\textrm{S/N}\left(\textrm{EW}\left(\textrm{H}\beta\right)\right) < 2$} for all 54 bins), implying that these bins are relatively faint.

Thus we are able to conclude that the potential impact of line emission from evolved stars is quite limited, and should not affect the metallicities we derive.

\subsubsection{Sample Properties}\label{sec:sample_properties}

\begin{figure*}
\includegraphics[width=\linewidth]{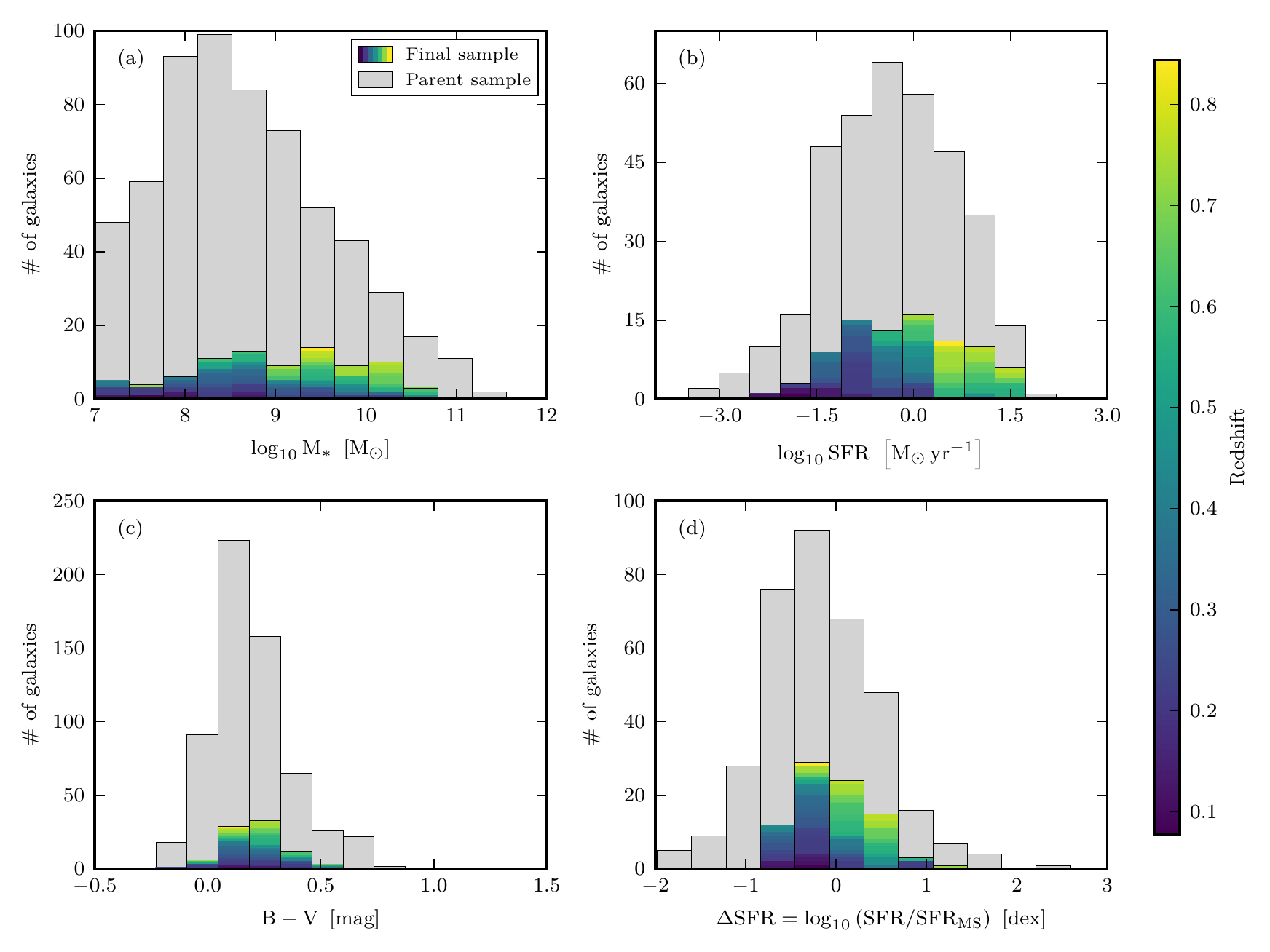}
\caption{The distributions of various global galaxy properties of our final sample are shown as coloured histograms.
The grey histograms show the distribution of galaxies from the parent sample.
Galaxies in the final sample are coloured by their redshift.
If the histogram bins were independent of redshift then each bin would be similarly coloured.
We show four properties: (a) stellar mass, (b) star-formation rate, (c) rest-frame Bessel B--V colour and (d) offset from the star-forming main sequence, accounting for redshift evolution.
Rest-frame colours are determined from the best-fit \textsc{magphys} spectrum to the photometry (see Section~\ref{sec:stellar_mass}).
While most galaxies in the parent sample have reliable photometry, allowing us to derive masses and colours, fewer galaxies in the parent sample have detected line-emission.
Consequently, there are fewer galaxies in the parent sample in panels (b~\&~d).
}
\label{fig:sample_description}
\end{figure*}

In Fig.~\ref{fig:sample_description} we present the global properties of our final sample of 84 galaxies.
Therein we show the distributions of stellar mass, SFR, rest-frame B--V colour and main-sequence offset ($\Delta\textrm{SFR}$)\footnote{We define $\Delta\textrm{SFR}$ to be the difference in the observed SFR relative to what would be expected for a galaxy on the main sequence, with the same stellar mass and at the same redshift. Here and throughout this paper we adopt the main sequence parametrization of \citet{2012ApJ...754L..29W}.}.
In addition, for comparison, we also plot the parent sample of all galaxies with a MUSE detected redshift.
This includes all 590 galaxies, even those that do not meet our selection criteria (Section~\ref{sec:selection_criteria}).

It can be seen that our final sample preferentially selects the more massive and more strongly star forming galaxies.
There is also a clear redshift dependence such that, in panel~(a), the low-mass galaxies are almost exclusively low-redshift galaxies.
The same is true for the SFR (panel~b), where the effect appears even stronger.

In contrast, both B--V colour and $\Delta\textrm{SFR}$ show different trends.
In panel~(c) we see that our final sample is fairly representative of the parent sample.
We note that the galaxy redshifts are relatively evenly distributed between each bin.

\begin{figure}
\includegraphics[width=\linewidth]{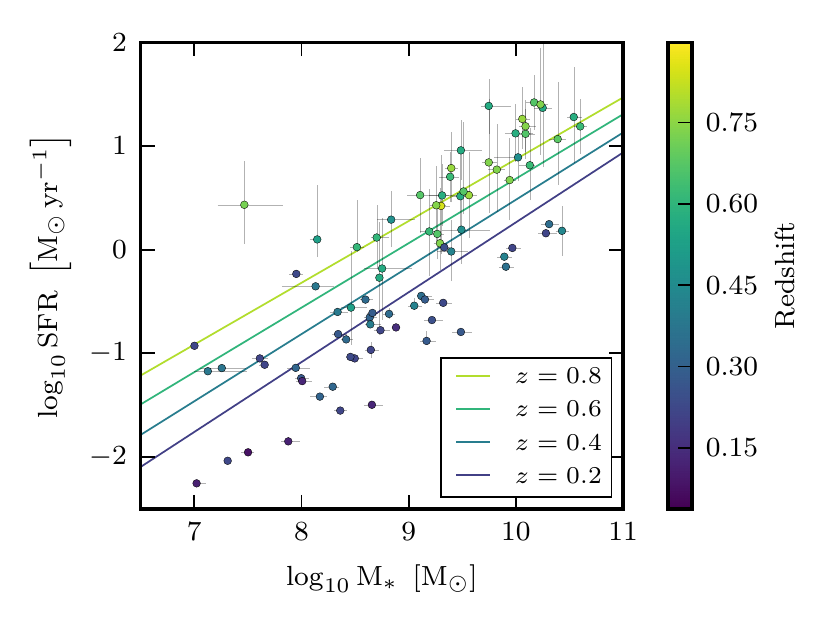}
\caption{Mass versus SFR for our final sample plotted as coloured circles.
For comparison, we display the main sequence at four different redshifts as solid lines, adopting the parametrization of \citet{2012ApJ...754L..29W}.
Note that the SFR of the low redshift galaxies are derived from \Halpha{} and \Hbeta{} lines, whilst those for the high redshift galaxies are derived from \Hbeta{} and faint \Hgamma{} lines.
Consequently the SFR errors are much smaller for the low redshift galaxies \mbox{($z\LA0.4$)}.
}
\label{fig:main_sequence}
\end{figure}

The same is true for the main-sequence offset parameter (panel~d) where, above $\Delta\textrm{SFR} \GA 0\ \textrm{dex}$, the final sample traces the shape as the parent sample and the redshifts are evenly distributed.
We can display this another way; in Fig.~\ref{fig:main_sequence} we show the mass--SFR correlation for our galaxies.
At high redshift our galaxies all lie above or on the main sequence.
And at low redshift a large fraction of galaxies are found below the main sequence.
However, note that the main sequence derived by \citet{2012ApJ...754L..29W} uses ultraviolet and infrared photometry to calculate their SFRs, whereas we use emission lines to derive our SFRs.
As a result the trend with redshift may not be purely a selection effect, since we could expect some systematic offsets and trends when comparing our SFRs relative to our adopted main sequence.

To summarize, at all epochs we are finding blue galaxies that lie on the upper half of the main sequence, however, at high redshift we are biased towards massive, strongly star-forming galaxies.

\section{Analysis}\label{sec:analysis}

Many of the galaxies we observe are heavily corrupted by seeing.
Furthermore, we must also aggregate (or ``bin'') spaxels\footnote{Spatial pixels.} together to increase the S/N of our data, creating additional resolution loss.
So, to recover the intrinsic metallicity gradient in our galaxies, we must model both the effect of seeing and binning on our data.

In \citetalias{2017MNRAS.468.2140C} we demonstrated such a method for inferring both the central metallicity, $\log_{10}Z_0$, and metallicity gradients, $\nabla_r\left(\log_{10}Z\right)$ in distant galaxies.
While the central metallicity and metallicity gradients inferred this way are inevitably model dependent, we performed a series of tests (including mock observations) to validate our procedure.
In this paper we apply this method to the sample of galaxies observed with MUSE.
We have made a few minor modifications to the method presented in \citetalias{2017MNRAS.468.2140C}.
For brevity here we only outline the method and highlight the changes.

In the next section we will first explain how we extract the emission-line fluxes (Section~\ref{sec:flux_extraction}).
We will then proceed to describe the fitting of our model to the data (Section~\ref{sec:model_fitting}).
Finally, we shall investigate how sensitive our recovered model parameters are to particular model inputs (Section~\ref{sec:sensitivity_analysis}).

\subsection{Emission line flux extraction}\label{sec:flux_extraction}

When extracting emission-line fluxes to measure metallicity gradients there is a trade off between the number of spatial bins and the S/N of the data within each bin.
One must choose a S/N threshold that is sufficiently high to minimize systematic errors in the emission-line measurements, whilst avoiding losing too much spatial information.

\subsubsection{Spatial Binning}\label{sec:spatial_binning}

In \citetalias{2017MNRAS.468.2140C} (appendix~C) we designed a binning algorithm that attempts to maximize the number of spatial bins above a S/N threshold.
In essence this algorithm performs successive passes over the data with an (initially small) fixed bin size.
When it is longer possible to find spatial bins above the S/N threshold, the bin size is then increased and the process repeated.
This continues up to a maximum allowed bin size, at which point we terminate the procedure and assign any remaining unbinned spaxels to a nearby bin.

When calculating the S/N of a spatial bin, we perform a full spectral fitting to the coadded spectrum.
For a successful bin, we require set emission lines to be detected at $\textrm{S/N} \geq 5$, where the set of tested emission lines is chosen on an object-by-object basis.
This set typically consists of four lines, the two highest-S/N Balmer lines and two highest-S/N forbidden lines.

There are some peculiarities that arise from our binning strategy.
To preserve as much radial information as possible we define our spatial bins in polar coordinates (in a plane inclined to the observer).
However, a direct consequence of working in a non-Cartesian coordinate system is that the pixels within a bin are not necessarily all close to each other in Cartesian space.
An additional oddity is caused by the successive passes with increasing bin size.
This can sometimes result in bins that are (partially or entirely) enclosed within another.
While neither of these effects are ideal, we remind the reader that we mirror the exact binning in the model (i.e. we give the binning segmentation map as a model input).
As a result, we view the strangeness of our binning scheme to be a fair trade for optimizing S/N whilst still preserving radial resolution.

Finally we note that, because we do not impose a minimum bin size, our spatial bins can be much smaller than the PSF.
Consequently emission-line fluxes of adjacent bins are not strictly statistically independent.
Since we do not incorporate such covariances in the model likelihood function, this might lead us to underestimate the errors in the derived model parameters.
However, in \citetalias{2017MNRAS.468.2140C} (section 3.1) we showed that our model nevertheless provides reasonable error estimates.
We therefore do not expect this issue to be of much concern.

\subsubsection{Spectral Fitting}\label{sec:spectral_fitting}

To extract the emission-line fluxes from the MUSE spectra we use the \textsc{platefit} spectral fitting code \citep{2004ApJ...613..898T,2004MNRAS.351.1151B}.
\textsc{platefit} applies a two-step process that first fits the stellar continuum (with emission lines masked) before fitting the nebular emission-line component (with the best-fit continuum subtracted).
Note that the procedure we employ here is identical to that presented in \citetalias{2017MNRAS.468.2140C} (section~4.2).
We summarize it briefly here.

The continuum fitting step of \textsc{platefit} does not fit either the redshift or velocity dispersion of the spectrum.
These two parameters must be provided in advance, and we do so as follows.
The redshift of the spectrum is obtained using \textsc{autoz} \citep{2014MNRAS.441.2440B}.
We wish this to be robust, so if the value determined by \textsc{autoz} deviates by more than $\pm500\,\mathrm{km}\,\mathrm{s}^{-1}$ from our initial redshift guess, then we default to that initial value.
To estimate the velocity dispersion we use \textsc{vdispfit}\footnote{\url{http://spectro.princeton.edu/idlspec2d_install.html}}.
At low S/N, however, \textsc{vdispfit} can yield outliers beyond the realistic range $\left[10 - 300\right]\,\mathrm{km}\,\mathrm{s}^{-1}$.
If values outside this range are produced, we adopt a default value of $80\,\mathrm{km}\,\mathrm{s}^{-1}$.

With the values of the redshift and velocity dispersion predetermined, the stellar continuum is fit using a combination of \citet{2003MNRAS.344.1000B} SPS model templates.
These templates form a grid of ten ages and four metallicites, spanning [5.2\,Myr, 11\,Gyr] and [0.004, 0.04], respectively.
The SPS models are built with the MILES stellar libray \citep{2006MNRAS.371..703S}, and therefore have a similar spectral resolution to that of our MUSE data (2.5\AA{}).
This resolution is sufficient for the purpose of obtaining a good stellar continuum subtraction.
If the stellar continuum is too faint / non-existent, the continuum fitting routine can fail.
In these cases we approximate the continuum by applying a running median filter to the spectrum (width 150\AA{}).

In the second \textsc{platefit} step, the best-fit continuum is subtracted from the observed spectrum.
The emission lines are modelled as Gaussian functions.
The velocity offset and velocity dispersions are the same for all emission lines.
However, unlike the continuum fitting, these two velocity components are free parameters and need not be specified in advance.

The emission-line fluxes are determined from the spectral fitting.
However, the formal emission-line flux errors are typically underestimated \citep[see][]{2013MNRAS.432.2112B}.
We rescale these formal errors using a S/N-dependent correction to obtain better estimates of the true flux error.
The correction factors were determined from duplicate Sloan Digital Sky Survey \citep[SDSS;][]{2000AJ....120.1579Y} observations.
Note that the SDSS spectra are of resolution comparable to MUSE, and the \textsc{platefit} correction factors are thus still applicable.

\subsection{Inferring metallicities}\label{sec:model_fitting}

We use a forward modelling method to correct for seeing effects allowing us to derive the true central metallicity and metallicity gradient of a galaxy.
With this paper, we are also publicly releasing the code.\footnote{\url{https://bitbucket.org/cartondj/metaldisc}}

In \citetalias{2017MNRAS.468.2140C} (section~2) we explained our method in detail and described how it can be used to fit the observed 2D emission-line flux distribution of galaxies.
Our method is almost identical to that presented therein, so we shall only briefly outline it here and highlight the changes.

We approximate a galaxy as an infinitesimally thin disc, inclined to the observer.
The disc is described by four fixed morphological parameters (RA, Dec, inc., PA).
Our model contains five free parameters: total SFR of the galaxy, $\textrm{SFR}_\mathrm{tot}$, central metallicity, $\log_{10}Z_0$, metallicity gradient $\nabla_r\left(\log_{10}Z\right)$, ionization parameter at solar metallicity, $\log_{10}U_{\sun}$, and the V-band optical depth, $\tau_V$.

The metallicity profile in the galaxy is assumed to be axisymmetric and is described by an exponential function
\begin{equation}\label{eq:metallicity_profile}
\log_{10}Z(r) = \nabla_r\left(\log_{10}Z\right) r + \log_{10}Z_0,
\end{equation}
where $r$ is the radius.
This parametrization is commonly used in high redshift studies \citep[e.g.][amongst many others]{2014MNRAS.443.2695S,2016ApJ...827...74W}.
However, as we discussed in the introduction, low redshift studies have suggested that the metallicity profile may be flat at both small ($\LA0.5\,r_e$) and large ($\GA1.5\,r_e$) radii \citep[see][]{2018A&A...609A.119S}.
That said, given the mass dependence they report for the inner flattening, and the fact that our galaxies are generally less massive than those presented therein, it is not obvious that we should expect inner flattening in our galaxies.
As for the outer flattening, we simply do not expect to be sensitive to the low S/N outskirts of our galaxies.
Therefore, in concordance with other high-redshift studies, we believe it is justified to adopt the simplistic parametrization of equation~\ref{eq:metallicity_profile}.
So note that, although we do not attempt to prove this to be the exact metallicity profiles of our galaxies, it is still a useful description for the overall variation in metallicity across a galaxy.

As per \citetalias{2017MNRAS.468.2140C}, we also assume the ionization parameter to be anti-correlated with metallicity
\begin{equation}\label{eq:ionization_parameter_coupling}
\log_{10}U\left(Z\right) = -0.8 \log_{10}\left(Z/Z_{\sun}\right) + \log_{10}U_{\sun},
\end{equation}
where $Z_{\sun}$ is the solar abundance and $\log_{10}U_{\sun}$ is the ionization parameter at the solar abundance.

To predict the observed emission-line ratios we use the photoionization models of \citet[][hereafter D13]{2013ApJS..208...10D}, who tabulate emission-line fluxes over a grid of metallicities and ionization parameters.
These models assume that most elemental abundances vary linearly with the oxygen abundance, except for carbon and nitrogen, whose dependence is empirically calibrated to observations of \ion{H}{ii} regions.
 
At each spatial position we interpolate the \citetalias{2013ApJS..208...10D} model grids to the appropriate values of metallicity and ionization parameter.
The modelled emission-line \emph{ratios} only depend on the radial coordinate; there is no azimuthal dependence.

We wish to include the \Hdelta{} and \Hepsilon{} emission lines in our model fit.
However, these Balmer lines are not provided by the \citetalias{2013ApJS..208...10D} photoionization models.
To include these lines we need to extend the \citetalias{2013ApJS..208...10D} photoionization models.
We do this by tabulating the Case B recombination ratios $j_{\textrm{H}\delta}/j_{\textrm{H}\beta}$ and $j_{\textrm{H}\epsilon}/j_{\textrm{H}\beta}$ as a function of $j_{\textrm{H}\gamma}/j_{\textrm{H}\beta}$.
By interpolating these at the \citetalias{2013ApJS..208...10D} photoionization model values of $L_{\textrm{H}\gamma}/L_{\textrm{H}\beta}$, we assign the appropriate $L_{\textrm{H}\delta}$ and $L_{\textrm{H}\epsilon}$ for each photoionization model.
The Case B recombination ratios were determined with \textsc{PyNeb} \citep{2015A&A...573A..42L} using atomic data from \citet{1995MNRAS.272...41S}.

\subsubsection{SFR Maps}

To model the emission-line \emph{luminosities} we need to model the SFR distribution in the galaxy.
For this one could adopt a parametric model, c.f. \citet{2016ApJ...827...74W} who assume an exponential disc (a disc where SFR declines exponentially with increasing radius).
However, since we have high-resolution HST imaging for all our galaxies, we prefer to relax this assumption and provide a 2D SFR map.
Unlike the emission-line \emph{ratios}, the modelled emission-line \emph{luminosities} do have an azimuthal dependence.

We assume that the distribution of stellar light in the HST images provides a rough approximation for the relative SFR distribution.
In the MUSE-Deep UDF fields we use the deep HST F775W imaging.
In all other fields we use the F814W band.
Although ideally we would like to use photometry that covers the rest-frame ultraviolet light in each galaxy, the necessary deep multi-wavelength HST imaging does not exist for all fields.
To strike a good balance we use the F775W and F814W filters, since they providing the S/N required.
This has the additional advantage that, because these filters have similar pivot wavelengths, we also ensure a degree of consistency between the fields.

We construct SFR maps of our galaxies by cropping the HST images to only include flux within an ellipse of radius $4 \times r_d$ along the major axis. 
This ellipse has the same morphology (RA, Dec, inc., PA) as above.
We inspect each image and alter the mask if necessary to ensure that we include all flux from the object and to remove other objects or defects.
When necessary we interpolate over these rather than mask them.
Negative flux values are set to zero.
The final result is an SFR map that represents the \emph{relative} spatial distribution of the SFR.
The absolute SFR values are determined by normalizing the map to the total SFR, $\textrm{SFR}_\mathrm{tot}$, which is a free parameter in the model.

From experience with our \citetalias{2017MNRAS.468.2140C} work, we found that it was quite important to include the complex spatial variations of the SFR in our model.
Using a SFR map (as opposed to simply assuming an exponential disc) appeared to improve the accuracy of our model.

However, including a SFR map does not correct for all effects of clumpy star-formation.
Indeed, a limitation of our model is that, by construction, we do not accommodate for galaxies with strong azimuthal metallicity variations\footnote{This limitation is by no means exclusive to our model. To properly treat poorly resolved galaxies, one would need a forward model with some non-parametric description of the 2D metallicity profile.}; we should anticipate that we may observe galaxies where the bright star-forming clumps have uncharacteristically low/high metallicities.

However, it should be noted that, for the low-redshift galaxies where azimuthal metallicity variations have been quantified, the variations are typically found to be quite small \citep[$\LA 0.05\,\textrm{dex}$; e.g.][]{2015A&A...573A.105S, 2016MNRAS.462.2715Z, 2017A&A...603A.113S}, and there is only the odd example of a galaxy with variations as large as $\approx 0.2\,\textrm{dex}$ \citep[e.g.][]{2017A&A...601A..61V}.

That said, even if there were large azimuthal variations in our galaxies, it does not mean that we cannot still make inferences in these cases because, in effect, we likely still measure the radial metallicity profile of brightest star-forming regions (i.e. a quasi flux-weighted metallicity profile).
Nevertheless, this would be in contrast with low-redshift observations, where we typically consider the measured metallicity profile to be describing the metallicity of the gas as a whole.
This distinction is an important one to bear in mind when interpreting metallicity gradients reported by all studies of poorly resolved high-redshift galaxies.

\subsubsection{Model Fitting}

With this galaxy model we are able to mimic the resolution loss due to the seeing and spatial binning.
Thus for every spatial bin we can generate a set of model fluxes that can be compared to those observed.

We fit emission lines that are observed at $\textrm{S/N} \geq 5$\footnote{We fit \forbidden{O}{ii}{3726}, \forbidden{O}{ii}{3729}, \forbidden{O}{ii}{3726,3729}, \forbidden{Ne}{iii}{3869}, \Hepsilon{}, \Hdelta{}, \Hgamma{}, \Hbeta{}, \forbidden{O}{iii}{4959}, \forbidden{O}{iii}{5007}, \forbidden{N}{ii}{6548}, \Halpha{}, \forbidden{N}{ii}{6584}, \forbidden{S}{ii}{6717}, \forbidden{S}{ii}{6731} and \forbidden{S}{ii}{6717,6731}. We exclude redundant lines, i.e. if \forbidden{O}{ii}{3726,3729} coadd is detected at $\textrm{S/N} \geq 5$, then we do not also fit \forbidden{O}{ii}{3726} and \forbidden{O}{ii}{3729}, even if they individually have a $\textrm{S/N} \geq 5$}.
For clarity, we emphasize that this threshold is applied for all detected emission-lines, not just the four chosen in Section~\ref{sec:selection_criteria}.
In other words, some emission lines may only be detected in a few bins, but a critical subset (two Balmer, two forbidden) will be detected in all bins.

\begin{table}
\caption{Priors on model parameters.
For each parameter we detail the type of prior and the range of values covered.}
\label{tab:priors}
\begin{tabular}{ccr@{\hspace{0.2em}}l}
\hline
Parameter & Prior type & \multicolumn{2}{c}{Range} \\
\hline
$\textrm{SFR}_\mathrm{tot}$ & Logarithmic & $\left[0.01, 100\right]$ & $\textrm{M}_{\sun{}}\textrm{yr}^{-1}$ \\
$\log_{10}Z_0$ & Linear & $\approx\left[-1.30, 0.70\right]$ & $\textrm{dex}$ \\
$\nabla_r\left(\log_{10}Z\right)$ & Linear & $\left[-0.3, 0.3\right]$ & $\textrm{dex}/\textrm{kpc}$ \\
$\log_{10}U_{\sun}$ & Linear & $\approx\left[-5.02, -1.42\right]$ & $\textrm{dex}$\\
$\tau_V$ & Linear & $\left[0, 4\right]$ & \\
\hline
\end{tabular}
\end{table}

To fit our model to the data we use the \textsc{MultiNest} multi-modal nested sampling algorithm \citep{2009MNRAS.398.1601F,2008MNRAS.384..449F,2013arXiv1306.2144F} that we access through a \textsc{Python} wrapper \citep{2014A&A...564A.125B}.
With \textsc{MultiNest} we can calculate the posterior probability distributions (posteriors) of our five model parameters.

The prior probability distributions (priors) that we place on our model parameters are outlined in Table~\ref{tab:priors}.
Except for two differences, these priors are mostly similar to the priors that were adopted in \citetalias{2017MNRAS.468.2140C}.

One difference is that we now adopt a logarithmic prior on $\textrm{SFR}_\mathrm{tot}$, where previously (for technical reasons) we had adopted a linear prior of $\left[0, 100\right]\,\textrm{M}_{\sun{}}\textrm{yr}^{-1}$.
For a normalization parameter we believe that a logarithmic prior is more appropriate.
Changing this prior should have little effect on the derived metallicity profiles, provided that the dust attention, $\tau_V$, is well constrained by the data.

A second and more significant change is that we now adopt a narrower prior on the metallicity gradient, $\nabla_r\left(\log_{10}Z\right)$ (previously we had adopted $\left[-0.5, 0.5\right]\,\textrm{dex}/\textrm{kpc}$).
In \citetalias{2017MNRAS.468.2140C} (appendix~B) we found that the inferred metallicity gradients could be sensitive to the choice of prior.
We attributed this sensitivity to the finite range of metallicities spanned by the photoionization model grid.
Metallicity profiles of galaxies with low central metallicities and steep negative metallicity gradients will bottom-out at the lowest metallicity allowed by the photoionization model grid.
As a direct result, negative metallicity gradients become somewhat degenerate in models with low central metallicities.
Likewise, positive metallicity gradients become degenerate in models with high central metallicities.

In \citetalias{2017MNRAS.468.2140C} we found that allowing an unnecessarily broad range of metallicity gradients could skew our inferences towards extreme values of the metallicity gradient.
Hence we have now chosen to limit our metallicity gradient prior to the range $\left[-0.3, 0.3\right]\,\textrm{dex}/\textrm{kpc}$.
This spans the range of currently measured metallicity gradients \citep[e.g.][]{2016A&A...585A..47M, 2016ApJ...827...74W, 2016ApJ...820...84L, 2017MNRAS.469..151B, 2018A&A...609A.119S}.
In the following section we will assess the sensitivity of our results to various parameters, including the width of this prior, and present a way to flag galaxies where the aforementioned degeneracy may be of concern.

\subsection{Sensitivity analysis}\label{sec:sensitivity_analysis}

With our model we perform a fit with five free parameters.
However, in our modelling we make particular choices, and fix certain parameters.
We should attempt to assess the sensitivity of our results to these, considering both the magnitude of systematic errors we could expect, and the significance of unaccounted random uncertainties.

Here we will investigate four possible sources of additional errors: the width of the metallicity gradient prior, a misestimation of the PSF, a miscalculation of the galaxy's inclination, and a misidentification of the galaxy centre.
To estimate the size of these effects we re-analyse our galaxies after altering one of the parameters (e.g. increasing the FWHM of the PSF by 10\%).
By comparing the difference between this alternate model and the fiducial model, we can assess the impact of unaccounted systematics.

\subsubsection{Sensitivity to the width of the metallicity gradient prior}\label{sec:sensitivity_prior}

\begin{figure*}
\includegraphics[width=\linewidth]{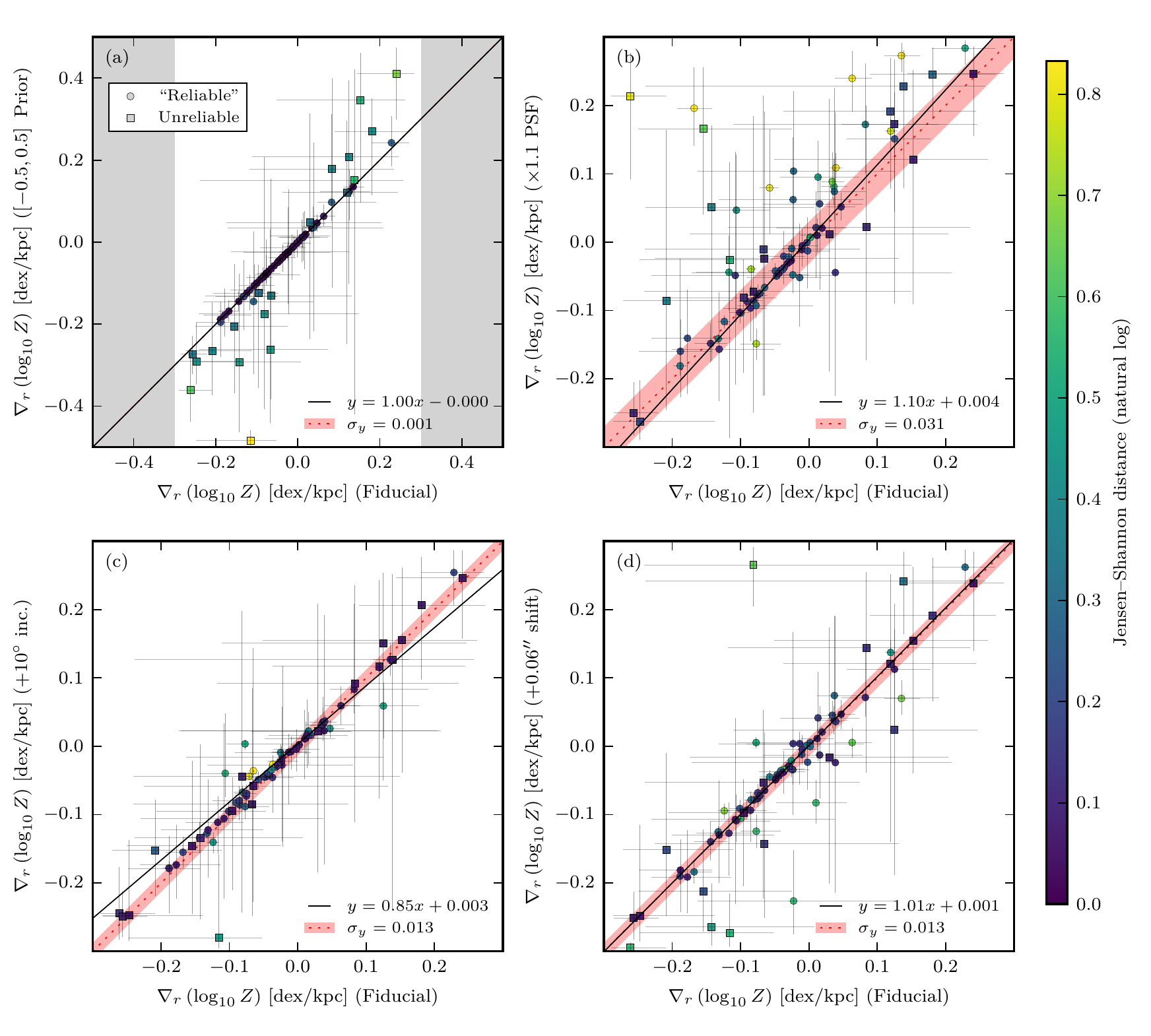}
\caption{Sensitivity analysis of model inputs.
On the horizontal axes we plot the metallicity gradients derived from our fiducial model.
On the vertical axes we compare the metallicity gradients derived from the same data, but with the alternate model.
The alternative models are identical to the fiducial model, except for one fixed parameter that has been adjusted slightly (see text for details).
A different alternative model is plotted in each of the four panels, showing the effects of: (a) the width of the metallicity gradient prior, (b) a misestimation of the PSF, (c) a miscalculation of the inclination and (d) misidentification of the galaxy centre.
The error bars show the $\pm1\sigma$ quantiles of the metallicity gradient posterior probability distribution.
N.B. the errors of the two models are not independent and are likely to be highly correlated, consequently the errors appear much larger than the scatter.
Each data point is coloured by the Jensen--Shannon distance between the 1D marginal posteriors of the fiducial and alternate models, highlighting the datapoints that deviate most from the 1:1 relation (shown as a red dotted line).
Squares indicate a subset of galaxies which have large Jensen--Shannon distances ($\textrm{JS}_\mathrm{dist} > 0.3$) when comparing the fiducial $[-0.3, 0.3]$ and the $[-0.5, 0.5]$ metallicity gradient priors.
We deem these galaxies as unreliable (see Section~\ref{sec:sensitivity_prior}).
In each panel a solid black line shows the best fit to the data, excluding unreliable galaxies and weighted by the inverse of the formal data errors.
A red shaded region indicates the (weighted) vertical scatter about this 1:1 relation, again excluding the unreliable galaxies.
Equations for the best fit and scatter are given in the bottom-right corner of each panel.
}
\label{fig:sensitivity}
\vspace{2em}
\end{figure*}

First we wish to identify the galaxies that are sensitive to the choice in metallicity gradient prior.
In Fig.~\ref{fig:sensitivity}(a) we compare the differences between two flat priors with different widths: $\left[-0.3, 0.3\right]$ (our fiducial prior) and $\left[-0.5, 0.5\right]\,\textrm{dex}/\textrm{kpc}$ (the alternate prior).
We see that most galaxies lie on the 1:1 line, indicating that they are robust against the choice of prior.
However, we note that some of the galaxies with steep metallicity gradients deviate significantly, showing a high degree of sensitivity to the prior.
This sensitivity to the prior indicates an unreliable estimate of the metallicity gradient.
These galaxies should be treated with care in any analysis.
Indeed, such behaviour might indicate that we have insufficient resolution (due to seeing and/or spatial binning effects) to constrain the metallicity profile of these galaxies.

To assess this sensitivity we need a metric to quantify the difference between the fiducial and alternate models.
For this we use the Jensen--Shannon distance applied to the metallicity gradient posteriors derived from the two models
\begin{equation}
\textrm{JS}_\mathrm{dist} = \sqrt{\frac{1}{2}\sum_i P_i \ln\frac{P_i}{M_i} + \frac{1}{2}\sum_i Q_i \ln\frac{Q_i}{M_i}},
\label{eq:jsdist}
\end{equation}
with
\begin{equation}
M_i = \frac{P_i + Q_i}{2},
\end{equation}
where $P_i$ and $Q_i$ are the discretized posteriors of the fiducial and alternate models, respectively.
We place an arbitrary cut on galaxies with large Jensen--Shannon distances \mbox{($\textrm{JS}_\mathrm{dist} > 0.3$)}.
From Fig.~\ref{fig:sensitivity}(a) this can be seen to be a good identifier of galaxies deviating from the 1:1 line.
In the remainder of this work we consider the derived metallicity gradients for these 19 galaxies to be suspect, and we will flag them as potentially unreliable.
For transparency we do not discard them and, unless otherwise stated, we include them in our statistics.

\subsubsection{Sensitivity to PSF errors}

The assumed PSF is perhaps one of the greatest sources of unaccounted error in our analysis.
In our model we assume to know the PSF perfectly, however, the true PSF is somewhat uncertain, particularly in the fields that do not contain bright stars.
Uncertainties in the PSF lead to additional random errors in our result.
More importantly, if we incorrectly parametrize the PSF, we could systematically bias our inferred metallicity gradients.

While it is challenging to truly characterize the effects of PSF errors, we can nevertheless attempt to estimate the magnitude of the problem.
We re-analyse our galaxies adopting two different PSFs, one where the FWHM is systematically 10\% smaller than the fiducial model, and another where it is 10\% larger.
We show the latter case in Fig.~\ref{fig:sensitivity}(b) and, as is to be expected, if we overestimate the PSF width then we will systematically overcorrect for seeing effects and infer systematically steeper metallicity gradients.

There is also noticeable scatter; a moderate uncertainty in the PSF will displace some galaxies significantly from the 1:1 line.
In general, (although not shown here) we find that the smallest galaxies are the most sensitive.
We can use the vertical displacement from the 1:1 line as an indication of the expected additional uncertainty.
We calculate the standard deviation (weighted by the inverse of the mean $1\,\sigma$ error), and find the additional uncertainty to be $\approx 0.026\,\textrm{dex}/\textrm{kpc}$, after excluding the unreliable galaxies\footnote{This estimate is the mean uncertainty obtained after averaging the two alternative models (one where the PSF is 10\% smaller and the other where the PSF is 10\% larger). Therefore this number differs slightly from the one presented in Fig.~\ref{fig:sensitivity}(b)}.
Naturally this value is only indicative and should not be treated as exact.
Not least because it will vary from galaxy to galaxy with the smaller galaxies being more affected.

\subsubsection{Sensitivity to galaxy inclination errors}

Another possible source of uncertainty is the inclination of the galaxy.
Whilst it is not so likely that we systematically miscalculate the inclination, there is certainly some random error.
In general the inclination is most uncertain for face-on galaxies, however, this is counteracted by the fact that the metallicity gradients derived from face-on galaxies are probably the most robust against errors in the inclination.

As a first guess we consider inclination uncertainties of $\pm10\degr{}$ and construct two alternate models.
We show one of these models in Fig.~\ref{fig:sensitivity}(c).
This plot indicates that if we overestimate the inclination, we will derive systematically smaller metallicity gradients.
However, we note that most galaxies have small Jensen--Shannon distances and clearly lie on the 1:1 line (with minimal scatter).
This suggests that the best-fit is in fact dominated by a few outliers, and therefore conclude that, in general, our model is robust against inclination uncertainties.
As before, from the vertical scatter we estimate the additional uncertainty to be $\approx 0.016\,\textrm{dex}/\textrm{kpc}$.

\subsubsection{Sensitivity to galaxy centre misidentification}

Finally, we address the impact of misidentifying the centre of the galaxy (the point about which the radial metallicity profile is defined).
We re-analyse the data, but shift the galaxy centre by $0.06''$ along the direction of the galaxy's major axis.
This distance is approximately one tenth the size of the MUSE PSF.
However, we remind the reader that the galaxy centre is actually defined from higher-resolution HST imaging.
While an absolute $0.06''$ shift will impact the smallest galaxies in our sample the most, it is arguably easier to define the centre of small compact galaxies than it is to define the centre of a large irregular galaxies.

If galaxies have a radial metallicity profile, we might na\"{i}vely expect that shifting away from the true centre would result in a flatter metallicity gradient.
However, in Fig.~\ref{fig:sensitivity}(d) we see no systematic trend towards flatter metallicity gradients.
Misidentifying the galaxy centre appears to add no systematic bias, but it could add moderate scatter to the inferred metallicity gradients.
We estimate this additional uncertainty to be $\approx 0.017\,\textrm{dex}/\textrm{kpc}$.

\subsubsection{Total additional uncertainty}\label{sec:total_uncertainty}

In summary we have considered three sources of additional uncertainty that are not incorporated by the model: misestimation of the PSF, miscalculation of the inclination and misidentification of the galaxy centre.
While, these are not necessarily the only contributions to the total additional uncertainty, they are certainly likely candidates.

\begin{table}
\caption{Estimated additional uncertainty in the metallicity gradient.
These are calculated for both with the unreliable galaxies excluded and also with them included.
Again we stress the approximate nature of these results and there may be other unaccounted sources of uncertainty.
}
\label{tab:additional_uncertainty}
\begin{tabular}{ccc}
\hline
Source of & Excluding & Including \\
uncertainty & unreliable galaxies & unreliable galaxies \\
 & [$\textrm{dex}/\textrm{kpc}$] & [$\textrm{dex}/\textrm{kpc}$] \\
\hline
PSF & $\approx0.026$ & $\approx0.034$ \\
Galaxy inc. & $\approx0.016$ & $\approx0.017$ \\
Galaxy centre & $\approx0.017$ & $\approx0.034$ \\
\hline
Total & $\approx0.035$ & $\approx0.050$ \\
\hline
\end{tabular}
\end{table}

In Table~\ref{tab:additional_uncertainty} we summarize our findings and add sources of uncertainty in quadrature to roughly estimate the total additional uncertainty.
When calculating the additional uncertainties we excluded the unreliable galaxies, so for full transparency we also present the estimated additional uncertainties if we also include the unreliable galaxies.
When considering additional uncertainties induced by misestimation of the PSF and misidentification of the galaxy centre, we note that by excluding the unreliable galaxies we decrease the additional uncertainty by about 23\% and 48\%, respectively.
Indeed it can be seen that some of the more discrepant galaxies in Fig.~\ref{fig:sensitivity}(b~\&~d) are also those flagged as unreliable (as defined via the Jensen--Shannon distance metric, see equation~\ref{eq:jsdist}). 
Thus it appears that a simple cut on galaxies that are sensitive to the prior, also partly helps to remove galaxies which are sensitive to other aspects of the model input.

\section{Results}\label{sec:results}

In this section we study the metallicity gradients of our galaxies and search for trends with galaxy size, mass and SFR.
Our main findings are as follows:

\begin{itemize}

\item The average galaxy in our sample has a negative metallicity gradient. 
However, there is considerable scatter and a few galaxies have positive metallicity gradients.

\item If we select only the largest galaxies ($r_d > 3\,\textrm{kpc}$), the scatter reduces.
There are, in fact, no large galaxies with inverted metallicity gradients.

\item We do not find significant trends between a galaxy's metallicity gradient and either its mass or its SFR.
The lack of correlation with the SFR is in contrast with other studies (as we shall discuss here and in \S\ref{sec:interpretation}).
\end{itemize}

From our MUSE observations we have constructed a sample of 84 galaxies between $0.08 < z < 0.84$.
At a $2\sigma$ significance level we identify 32 galaxies with negative metallicity gradients and 7 with positive gradients.
We classify 26 galaxies as having metallicity gradients that are \emph{not inconsistent} with zero.
Hereafter we shall term these galaxies as having flat metallicity gradients, a name which should not be over-interpreted; we cannot (and do not) claim galaxies with flat metallicity gradients to have no metallicity gradient whatsoever, but rather our data/analysis lacks the statistical power to say otherwise.
A further 19 galaxies are flagged as unreliable for being sensitive to the metallicity gradient prior (Section~\ref{sec:sensitivity_prior}).
In Table~\ref{tab:results} we provide the inferred central metallicities and metallicity gradients of our galaxies.
Additionally, we provide an image atlas of the galaxies in Fig.~\ref{fig:stamps}.
(As electronic-only supplementary material we provide plots showing both the model fit quality and parameter constraints. Examples of each are given in Figs.~\ref{fig:model_fit}~\&~\ref{fig:triangle}, respectively)

Our first result is that on average our galaxies have a negative metallicity gradient with a median of $-0.039^{+0.007}_{-0.009}\, \textrm{dex}/\textrm{kpc}$ (using bootstrap and Monte Carlo resampling).
We also apply a Wilcoxon signed-rank test and reject the null hypothesis that the average metallicity gradient is zero ($p=0.0014$).
Contrast this with the findings of \citet{2016ApJ...827...74W} who, in a study of $0.6 \LA z \LA 2.6$ galaxies, find the average metallicity gradient to be flat.

\begin{figure}
\includegraphics[width=\linewidth]{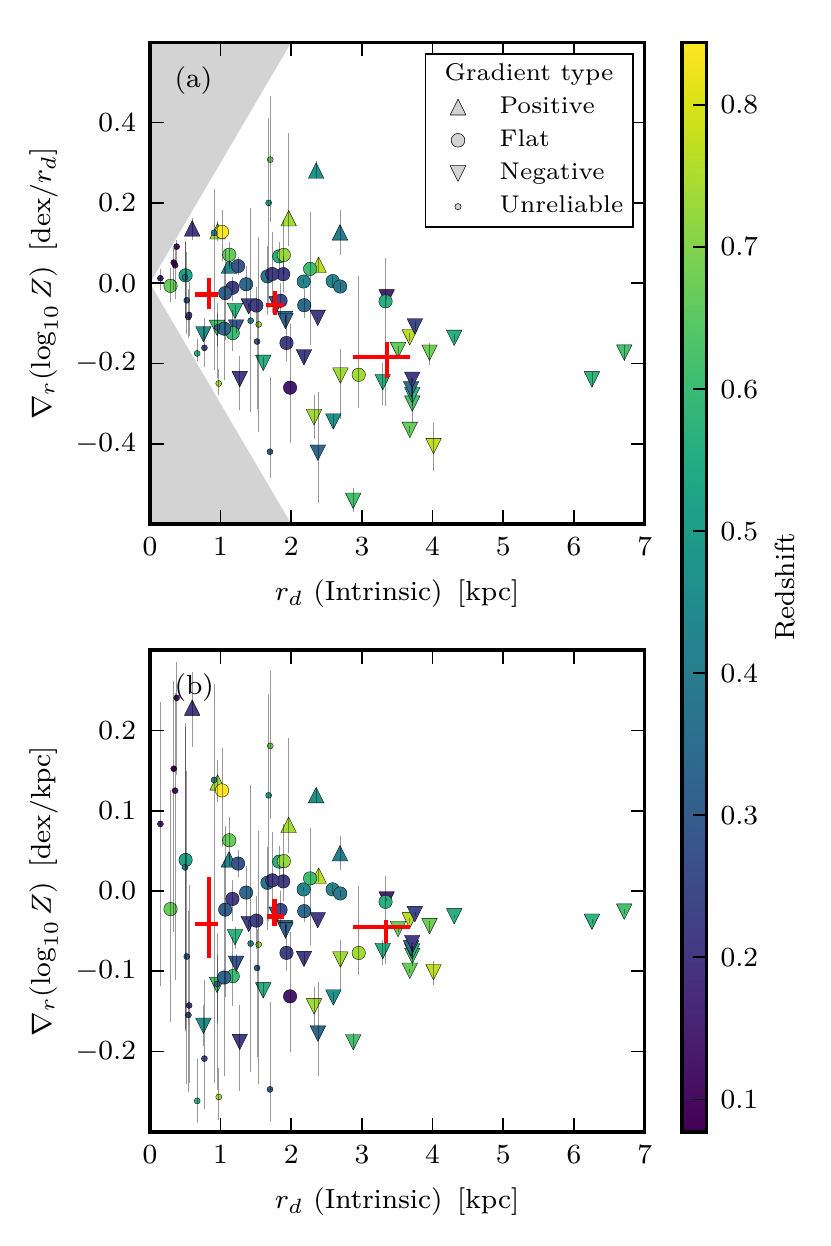}
\caption{Metallicity gradients of galaxies as a function of their size.
On the horizontal axes we plot the disc scale-length of the galaxies.
On the vertical axes we show the metallicity gradient.
In panel (a) we have normalized the metallicity gradient to the disc scale length, while in (b) we display the same data but with the metallicity gradient expressed in physical units.
Symbols indicate our metallicity gradient classification: triangles pointing up/down are galaxies with $2\sigma$ positive/negative gradients, large circles represent galaxies that have metallicity gradients consistent we being flat, and small circles indicate galaxies flagged for their sensitivity to the prior (the squares in Fig.~\ref{fig:sensitivity}(a)).
Data points are coloured according to the galaxy's redshift.
We overplot three red crosses, which indicate the median trend of the metallicity gradient with size.
Errors on the median are determined by bootstrapping the sample, and Monte Carlo sampling of the errors (this includes unreliable galaxies).
In panel (a), a grey shading denotes the region disallowed by our prior on the metallicity gradient.}
\label{fig:grad_vs_size}
\end{figure}

In Fig.~\ref{fig:grad_vs_size} we present the metallicity gradients of galaxies as a function of their size.
Panels~(a~\&~b) show the metallicity gradient expressed in scaled units ($\textrm{dex}/r_d$) and physical units ($\textrm{dex}/\textrm{kpc}$), respectively.
Throughout the remainder of this work we will present metallicity gradients in scaled units that normalize for the galaxy's size.
At low redshift there is some consensus that when expressed this way, isolated massive galaxies ($\ga 10^8\,\textrm{M}_{\sun{}}$) share a common value for the metallicity gradient \citep{2014A&A...563A..49S,2015MNRAS.448.2030H}.
While this may not be true at higher redshift, we will nonetheless use scaled units.

From Fig.~\ref{fig:grad_vs_size} it is clear that the average galaxy has a negative metallicity gradient.
However, it is also clear that there is considerable scatter.
There is a noticeable trend for the median metallicity gradient to become more negative with increasing galaxy size.
We see that amongst the large galaxies ($\GA 3\,\textrm{kpc}$) there are no galaxies with positive metallicity gradients, regardless of redshift.
Indeed almost all large galaxies have negative metallicity gradients.
Conversely, the small galaxies ($\LA 3\,\textrm{kpc}$) present a range of metallicity gradients, some negative and some positive.

\begin{table}
\caption{Analysis of unexplained (intrinsic) scatter in the metallicity gradients.
We model the metallicity gradients as if they were normally distributed with mean, $\mu$, and standard deviation, $\sigma_\mathrm{int}$.
The galaxies are divided into two groups: those smaller, and those larger than $r_d = 3\,\textrm{kpc}$.
We show results for metallicity gradients expressed in both physical units ($\textrm{dex}/\textrm{kpc}$) and scaled units ($\textrm{dex}/r_d$).
The unreliable galaxies, which are sensitive to the prior, are not included in this analysis.
Including them would not significantly alter the results.
}
\label{tab:intrinsic_scatter}
\begin{tabular}{ccccc}
\hline
Gradient  & \multirow{2}{*}{Size} & \multirow{2}{*}{$\mu$} & \multirow{2}{*}{$\sigma_\mathrm{int}$} & \# of \\
units & & & & galaxies\\
\hline
\multirow{2}[3]{*}{$\textrm{dex}/\textrm{kpc}$} & $\le 3\,\textrm{kpc}$ & $-0.027^{+0.013}_{-0.014}$ & $0.083^{+0.011}_{-0.010}$ & 49 \bigstrut[t] \\
 & $> 3\,\textrm{kpc}$ & $-0.054^{+0.008}_{-0.008}$ & $0.030^{+0.008}_{-0.006}$ & 16 \bigstrut \\
\hline
\multirow{2}[3]{*}{$\textrm{dex}/r_d$} & $\le 3\,\textrm{kpc}$ & $-0.045^{+0.020}_{-0.020}$ & $0.128^{+0.020}_{-0.015}$ & 49 \bigstrut[t] \\
& $> 3\,\textrm{kpc}$ & $-0.212^{+0.028}_{-0.027}$ & $0.105^{+0.028}_{-0.020}$ & 16 \bigstrut \\
\hline
\end{tabular}
\end{table}

The scatter in the metallicity gradient appears to sharply increase below $\LA 3\,\textrm{kpc}$.
However, we see that the error bars of the small galaxies also increase.
It is therefore important to ask whether there is a true increase in the \emph{intrinsic} scatter in the small galaxies.
In Table~\ref{tab:intrinsic_scatter} we present an analysis of the amount of intrinsic scatter in both the small and large galaxies.
For this we model the metallicity gradients as normally distributed with mean, $\mu$, and standard deviation, $\sigma_\mathrm{int}$.
This analysis accounts for the full posterior distributions on the metallicity gradient.
We see that there is indeed an increase in the intrinsic scatter in the small galaxies ($\LA 3\,\textrm{kpc}$) and that the mean is overall negative.

However, we caution that although we call this intrinsic scatter, it is perhaps more honest to call it unexplained scatter.
It could simply be that our model is underestimating the true error in the metallicity gradient.
Indeed it is somewhat suspicious that we see an increase in the scatter towards the smaller galaxies.
We na\"{i}vely would expect small galaxies to be more sensitive to model specification errors, i.e. errors in the PSF or inclination (see Section~\ref{sec:sensitivity_analysis}).
Nevertheless, the sharpness of the increase in scatter for galaxies with sizes $\LA 3\,\textrm{kpc}$ is certainly worthy of note.

\begin{figure}
\includegraphics[width=\linewidth]{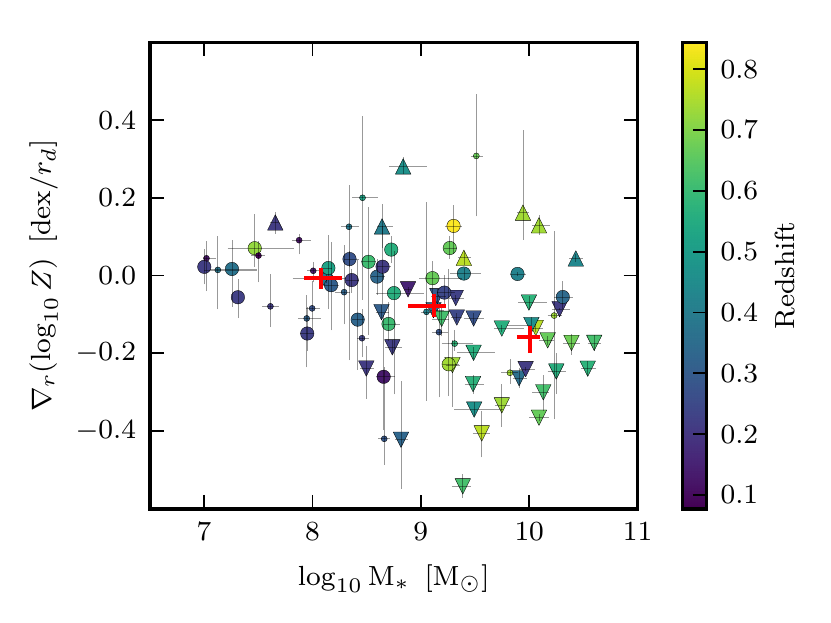}
\caption{Metallicity gradients of galaxies as a function of their stellar mass.
See Fig.~\ref{fig:grad_vs_size} for a description of the symbols and colours.
}
\label{fig:grad_vs_mass}
\end{figure}

It is well known than there exists a correlation between the size and total stellar mass of a galaxy \citep[see][and references therein]{2014ApJ...788...28V}.
So, given that the galaxies with positive metallicity gradients are exclusively small, we also might expect them to be also less massive.
However, this does not appear to be the case in Fig.~\ref{fig:grad_vs_mass}, where we plot the metallicity gradients against the total stellar mass of the galaxy.
Galaxies with positive metallicity gradients span a similar mass range as those with negative gradients.
That said, it does appear that on average the metallicity gradient decreases with increasing mass.
This trend would be qualitatively similar to that found in studies of low-redshift galaxies, e.g. \citet{2017MNRAS.469..151B} and (somewhat indirectly) the works of \cite{2013ApJ...764L...1P} and \cite{2016MNRAS.463.2799I}.
However, because of our galaxy selection, it is hard to interpret the trend we observe.
With a larger sample one might be able to disentangle the coupled selection effects on mass, size and redshift.

It is easier, however, to discuss the more massive portion of our sample.
For example, at masses above $\approx10^{9}\,\textrm{M}_{\sun{}}$ we notice there is a large range of metallicity gradients, with hints that the scatter is perhaps greater in the higher-redshift galaxies.
This certainly is in direct contrast to the low-redshift ($z \ll 0.1$) literature result of a common metallicity gradient.
This discrepancy might be resolved if many of these galaxies are not isolated, but are instead interacting with other galaxies.
Because, interacting galaxies typically have flatter metallicity gradients than would otherwise be expected \citep{2012ApJ...753....5R}.

\begin{figure}
\includegraphics[width=\linewidth]{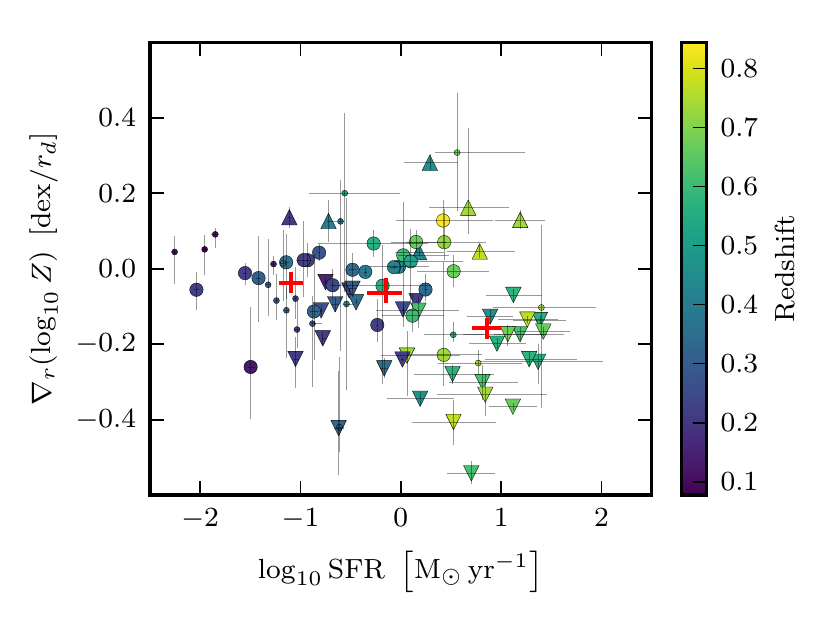}
\caption{Metallicity gradients of galaxies as a function of their star formation rate.
See Fig.~\ref{fig:grad_vs_size} for a description of the symbols and colours.
}
\label{fig:grad_vs_sfr}
\end{figure}

Since interactions could trigger increased star-formation rates in interacting galaxies, \citet{2014MNRAS.443.2695S} suggested that the scatter in the observed metallicity gradients could be explained by the SFR.
In Fig.~\ref{fig:grad_vs_sfr} we compare the metallicity gradient to the SFR, however, we do not observe a trend towards more positive metallicity at high SFRs.
Admittedly the association between interacting galaxies and higher SFR might be weak.
However, in fact we see a slight trend in the opposite direction with the metallicity gradient becoming more negative at high SFRs.
We caution that this trend could be a manifestation of our sample selection (i.e. because our sample is incomplete).
Notably the SFR of a galaxy in our sample correlates with its redshift.
This is in part because at higher redshifts we are biased towards the brighter, more strongly star-forming galaxies.
However, it is also in part because we expect the galaxy main-sequence to be evolving (the average SFR is less today than it was at $z \approx 1$).

\begin{figure}
\includegraphics[width=\linewidth]{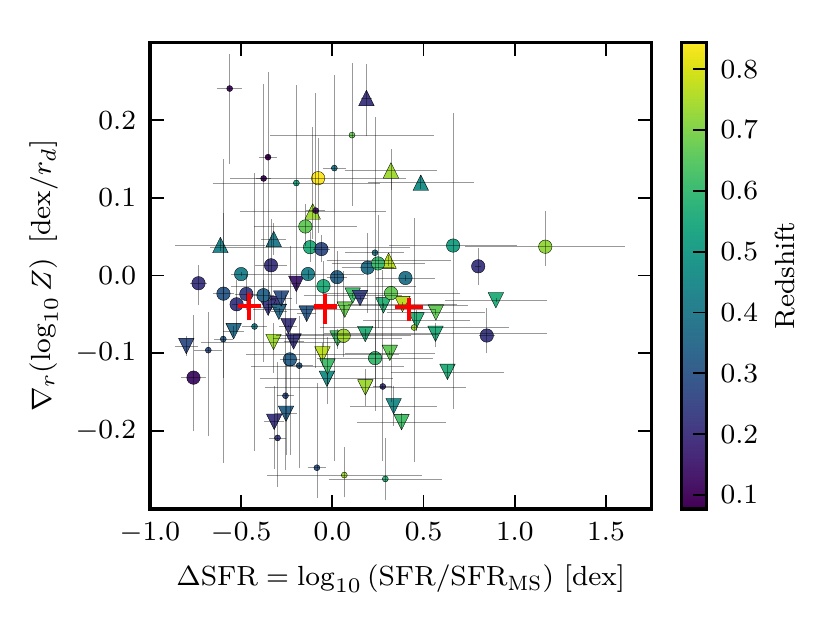}
\caption{Metallicity gradients of galaxies as a function of $\Delta\textrm{SFR}$, the SFR of a galaxy relative to what is expected for a galaxy on the main-sequence with identical mass and redshift.
We use the main-sequence parametrization of \citet{2012ApJ...754L..29W}.
See Fig.~\ref{fig:grad_vs_size} for a description of the symbols and colours.
}
\label{fig:grad_vs_dsfr}
\end{figure}

So to put the galaxies on an even footing, we normalize the SFR relative to the main sequence.
We define the main-sequence offset, $\Delta\textrm{SFR}$, as the difference between a galaxy's SFR and the SFR of a galaxy on the main sequence which has identical mass and redshift, adopting the parametrisation of the main sequence evolution from \citet{2012ApJ...754L..29W}.
In Fig.~\ref{fig:grad_vs_dsfr} we show the metallicity gradients of our galaxies against $\Delta\textrm{SFR}$.
We see that the trend towards negative gradients with increasing SFR has now disappeared.
In fact, this projection shows no trend whatsoever; the average gradient is more or less constant, and there is large scatter irrespective of a galaxy's position relative to the main-sequence.

This is in contrast to the results of \citet{2014MNRAS.443.2695S} who find a positive correlation between specific star-formation rate (sSFR) and metallicity gradient.
They find galaxies that are vigorously forming stars have flatter or even positive gradients.
The authors suggest that an event causing infall of metal-poor gas would simultaneously trigger intense star formation and reduce the central metallicity (thus flattening or inverting the metallicity profile).
Our findings challenge this explanation for the cause of inverted metallicity gradients.
However, our results do not automatically preclude the mechanisms of galaxy--galaxy interactions and/or cold flows for triggering star formation.

In the next section we shall discuss our results in more detail, placing them in context with other observations and theoretical work.

\section{Discussion}\label{sec:discussion}

\subsection{Literature comparison}

\begin{figure*}
\includegraphics[width=\linewidth]{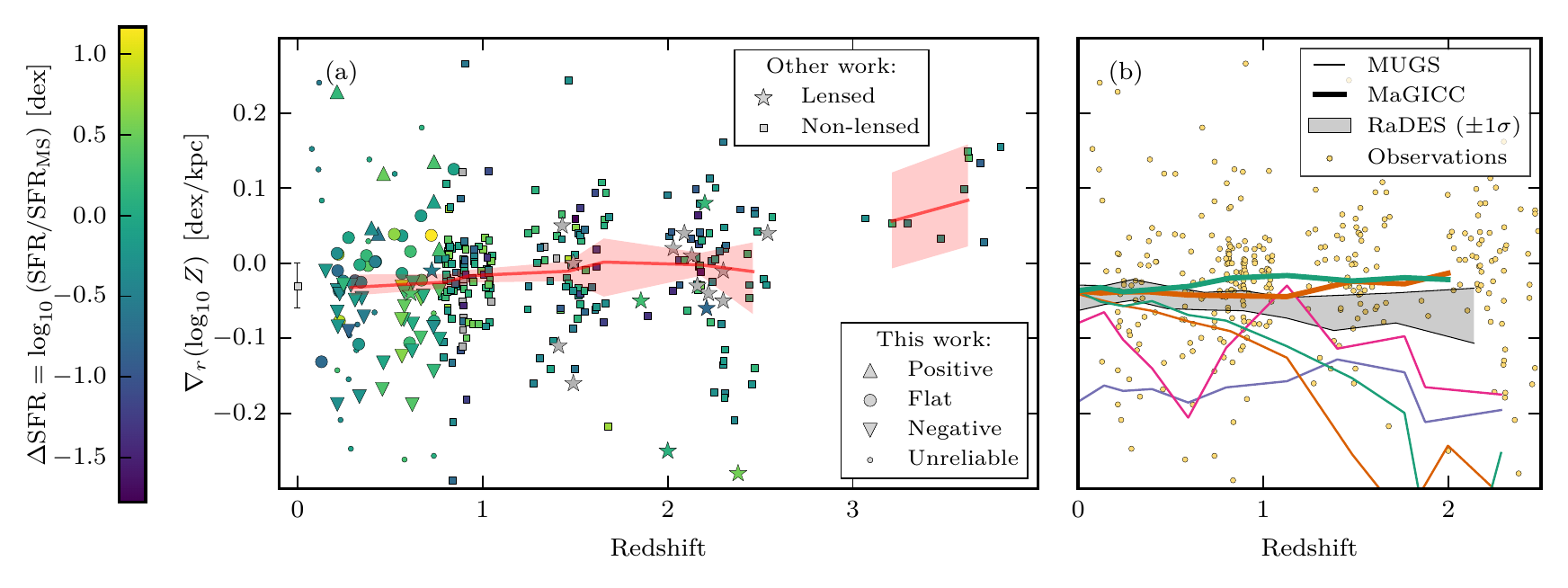}
\caption{Comparison of our results with the literature, including both observations and numerical simulations.
(Left) We compare our observational results against other studies.
These metallicity gradients come from studies of galaxies that have been gravitationally lensed (star symbols), and non-lensed galaxies (squares).
Where possible we colour data points to indicate the SFR relative to the main sequence ($\Delta\textrm{SFR}$).
The $z\approx0$ data point represents the mean and scatter in metallicity gradients reported in \citet{2014A&A...563A..49S}.
We compile lensing results from \citet{2011ApJ...732L..14Y}, \citet{2013ApJ...765...48J}, \citet{2015AJ....149..107J}, \citet{2016ApJ...820...84L}, \citet{2017ApJ...837...89W} and Patr\'{i}cio et al. (in prep.).
The non-lensed results are gathered from \citet{2012A&A...539A..93Q}, \citet{2012MNRAS.426..935S}, \citet{2014MNRAS.443.2695S}, \citet{2014A&A...563A..58T} and \citet{2016ApJ...827...74W}.
We also overplot a red trend line which indicates the median metallicity gradient in equally spaced redshift bins (excluding both the unreliable and $z\approx0$ galaxies).
The shaded region around this line represents the $1\sigma$ scatter of the median, derived from a bootstrap and Monte Carlo resampling of the errors.
(Right) Observational results (ours and others) compared with various numerical simulations.
The grey shaded region indicates the $\pm 1\sigma$ spread of metallicity gradients from 19 RaDES simulated galaxies \citep{2012A&A...547A..63F}.
Thin coloured lines indicate four individual MUGS simulated galaxies \citep{2010MNRAS.408..812S}.
Two of these galaxies (the orange and green lines) were re-simulated within the MaGGIC simulations \citep{2013A&A...554A..47G} and are shown as thick orange/green lines.
}
\label{fig:litsim_comparison}
\end{figure*}

The work presented here is the first large systematic study of metallicity gradients in galaxies between $0.1 \LA z \LA 0.8$.
However, there have been numerous studies of metallicity gradients in galaxies at lower and higher redshifts.
In Fig.~\ref{fig:litsim_comparison}(a) we plot our results alongside several of these studies.
Many of the high-redshift observations have used IFS techniques similar to ours, although some use observations of gravitationally lensed galaxies.
Ideally these lensed observations will have sufficient resolution such that seeing does not significantly affect the observed metallicity gradient.
Therefore it is worth noting that the lensed and non-lensed studies are consistent with one another, with both having means close to zero.
Furthermore, apart from some lensed galaxies with very steep negative metallicity gradients, the scatter in the two distributions is similar.

There is, however, a slight discontinuity between our observations and the other studies at $0.8 \LA z \LA 2.6$.
This transition is perhaps most visible in the increased scatter of our observations.
But there is also a shift in the average gradient to slightly negative gradients in our sample.
While these effects could be attributed to the real evolution of metallicity gradients with cosmic time, a discontinuity would argue for less astrophysical causes.

There are a few plausible explanations, of which the most concerning for us is that this could highlight an issue with our method for determining metallicity gradients.
A systematic overcompensation for seeing effects certainly could produce steeper gradients.
It is harder, however, to conceive of systematic effects that would produce a shift away from a non-zero average gradient.
In support of our results, a slightly negative average gradient is in fact consistent with observations in the low-redshift Universe \citep[e.g.][]{2014A&A...563A..49S}.
In particular, \citet{2016A&A...595A..62P} report positive metallicity gradients occurring in 10\% of their $z\sim0.02$ sample.
This is similar to the $\left(8\pm3\right)\!\%$ we find here (7 out of 84 galaxies).
Note, given the $2\sigma$ significance level that we choose when classifying galaxies as having positive metallicity gradients, we estimate a non-negligible number of false positives (0.77 galaxies)\footnote{$(26 + 7) \times 2.275\%$ -- The number of galaxies with flat or positive metallicity gradients, multiplied by the one-tailed $2\sigma$ probability.}.

There is another potential reason for the discontinuity between the observational studies.
The vast majority the $0.8 \LA z \LA 2.6$ galaxies have metallicity gradients determined using the N2 ratio ($\forbidden{N}{ii}{6584} / \Halpha{}$) and the calibration proposed by \citet[][hereafter PP04]{2004MNRAS.348L..59P}.
Whereas, we use a very different method for deriving metallicity.
It is well known that there are large discrepancies between different metallicity determination methods \citep[see][]{2008ApJ...681.1183K}.
So it is worth noting, although not demonstrated here, that combining the \citetalias{2013ApJS..208...10D} photoionization models with our metallicity and ionization parameter coupling (equation~\ref{eq:ionization_parameter_coupling}) tends to provide similar metallicity gradients to those produced by the method of \cite[][hereafter M08]{2008A&A...488..463M}.
The calibration adopted by \citetalias{2008A&A...488..463M} has a shallower dependence of the N2 ratio on metallicity than that used by \citetalias{2004MNRAS.348L..59P}.
Therefore, using \citetalias{2008A&A...488..463M} yields systematically steeper metallicity gradients than \citetalias{2004MNRAS.348L..59P}.
And, consequently, when comparing our work with the other studies that use the \citetalias{2004MNRAS.348L..59P} calibration, one might also expect our metallicity gradients to be systematically steeper.


An advantage of our method is that it produces a self-consistent metallicity gradient analysis, independent of the available emission-lines.
With our MUSE observations we lose emission lines redward of \Halpha{} at $z \GA 0.4$.
However, we do not observe a systematic shift in our measurements at $z \approx 0.4$, suggesting that our method is indeed self-consistent.

On a related note, it is questionable whether any of these metallicity methods are valid at high redshift, since most metallicity calibrations and photoionization models are designed for low-redshift interstellar medium (ISM) conditions.
For example, it is generally accepted that the electron density of the ISM was higher at earlier times \citep{2014ApJ...787..120S,2016ApJ...816...23S}, but the \citetalias{2013ApJS..208...10D} photoionization models we use are computed only at low densities ($\approx 10\,\textrm{cm}^{-3}$).
At earlier times the ionization parameter, the hardness of the ionizing spectrum, or the nitrogen-to-oxygen abundance ratio may also have been different, although there is little consensus on this \citep[e.g.][and references therein]{2016ApJ...822...42O, 2017ApJ...835...88K, 2016ApJ...826..159S}.
Our method is certainly not immune to these issues.
However, since our observations are all below $z \approx 0.8$ it may not be inappropriate to apply the same assumptions as we use at low redshift.
It is worth noting that because we marginalize over galaxy wide variations in the ionization parameter, our method may partially mitigate against some of the variations in ISM conditions.
This is because (to first order) the largest variations in nebular emission-line spectra are typically due to metallicity and ionization parameter \citep{2000ApJ...542..224D}.

\subsection{Interpretation}\label{sec:interpretation}

In the work presented here we find that, on average, galaxies between $0.1 \LA z \LA 0.8$ have a negative metallicity gradient.
However, there is considerable scatter, with some galaxies exhibiting a positive metallicity gradient.
As reported by \citet{2012A&A...540A..56P}, numerical simulations predict that there is intrinsic scatter in the metallicity gradient.
This can be seen in Fig.~\ref{fig:litsim_comparison}(b) when comparing the scatter \emph{within} the MUGS galaxies \citep{2010MNRAS.408..812S} and within the RaDES galaxies \citep{2012A&A...547A..63F}.
One should note that the difference \emph{between} the MUGS and RaDES simulations are, however, attributable to differences in the numerical recipes for star formation and feedback in the various simulations.
Flatter metallicity gradients arise from the model prescriptions with more intense feedback.
For example, the MaGICC simulations \citep{2013A&A...554A..47G} re-simulate two of the MUGS galaxies using an enhanced feedback recipe.
The increased feedback produces galaxies with metallicity gradients that are consistently flatter at all redshifts.
\citet{2013A&A...554A..47G} attribute this to central gas that is lost in outflows and later recycled back into the galaxy, but at larger radii.

The intrinsic diversity in metallicity gradients that we observe (and which is also present in the simulations) would indicate that at earlier times there is no common metallicity gradient.
It is perhaps surprising that we observe a common abundance gradient in the Universe today.
That said, as we shall later suggest, this contradiction is not necessarily as acute as it would first appear.
A common metallicity gradient may only exist in large galaxies.

Even though these simulations can produce a large range of metallicity gradients, they typically do not reproduce the same inverted metallicity gradients that we and others observe.
The simulations tend to produce galaxies with negative metallicity gradients (both steep and shallow).
Some of these simulations have next to no redshift dependence, while others suggest that the metallicity gradient was steeper at earlier times.
Despite this variety, the simulations never produce galaxies with positive metallicity gradients.
Indeed, with the classical understanding of inside-out growth one expects negative metallicity gradients \citep{1999A&A...350..827P}.
Radial mixing of gas could flatten the metallicity gradient, but it is hard to conceive of secular processes that could produce positive metallicity gradients.

Consistent with other higher redshift studies, we identify some of our galaxies that have significantly positive metallicity gradients.
Contrast this with galaxies at much higher redshift ($z \approx 3.4$), which all have centres that are systematically more metal poor than their outskirts \citep{2014A&A...563A..58T}.
In related work, \citet{2010Natur.467..811C} attribute this metallicity gradient inversion to cold flows \citep[e.g.][]{2009Natur.457..451D}.
Cold flows are streams of cold gas that penetrate the hot galaxy halo and fuel star formation in galaxies.
It is argued that if this metal-poor cold gas can reach the innermost regions of a galaxy (the deepest part of the gravitational potential) then this would explain the metallicity gradient inversion.

It should be noted that cold flows are not the only way to transport metal-poor gas to the inner portions of a galaxy.
As suggested by \citet{2012ApJ...753....5R}, galaxy--galaxy interactions could cause radial inflow of gas though the galaxy's disc.
Metal-poor gas from the outskirts is therefore deposited in the galaxy centre.
This is supported by simulations, which find that interacting galaxies have flatter (although never positive) metallicity gradients \citep{2010ApJ...710L.156R,2012ApJ...746..108T,2017MNRAS.472.4404S}.
However, work by \citet[][Ch.~8]{divoy:tel-01158668} suggested that merging systems can perhaps exhibit positive metallicity gradients once simulations have been downgraded to mimic the effects of spatial resolution loss and/or noise.
Nevertheless, it is not clear whether galaxy interactions can truly invert the metallicity gradient.

In contrast to the inflow mechanisms above, outflows could also provide an alternative explanation for positive metallicity gradients.
Intense centrally-concentrated star formation could produce significant outflows that entrain metal-rich gas.
Using simple analytical chemical evolution arguments, \citet{2010Natur.467..811C} disfavoured a wind scenario as it would require mass outflow rates far in excess of the observed SFR.
That said, \citet{2014A&A...563A..58T} point out that if this metal-rich gas falls back preferentially onto the outer regions of the galaxy, it could raise the outer metallicity.
This ``fountaining'' could then enhance the ability of winds to produce positive metallicity gradients and hence reduce the required outflow rates down to more realistic levels.

The inflow and outflow scenarios need not be mutually exclusive.
Gas accretion could trigger intense star formation that drives the winds.
In either case (inflows and/or outflows) we should expect to observe elevated star-formation rates.
In support of this \citet{2014MNRAS.443.2695S} and \citet{2016ApJ...827...74W} find weak correlations between the specific star-formation rate (sSFR) and metallicity gradient (galaxies with more intense star-formation have flatter or even positive metallicity gradients).
In contrast, we do not find systematically different metallicity gradients in galaxies with elevated SFR (Fig.~\ref{fig:grad_vs_dsfr}).
A difference in the sample selection might plausibly account for the difference between our results and previous studies.
Alternatively, one should also consider that if the intense star-formation is confined to a galaxy's centre, the \emph{global} SFR of a galaxy may not be a very sensitive indicator of inflow or outflow events.
Furthermore, the timescale for observing an inverted metallicity gradient might be different from that to observe an elevated SFR.
As such, some of the galaxies observed with elevated SFRs are likely moving vertically down the main sequence (i.e. have a declining SFR at the instant we observe them).
Therefore timescale differences might further weaken any correlation between the specific star-formation rate and metallicity gradient.

There is perhaps another way to reconcile the lack of correlation between metallicity gradient and star-formation intensity.
Contrary to our earlier assertion, it has been recently proposed by \citet{2017MNRAS.467.1154S} that inverted metallicity gradients can actually arise mostly within the framework of secular inside-out growth.
This is achieved in conjunction with outflows that recycle enriched gas from the galaxy centre, transporting it to larger radii.
\citet{2017MNRAS.467.1154S}, however, show that these positive metallicity gradients do not persist throughout a galaxy's life, and in fact are only expected to exist at early times in a galaxy's evolution.
Unfortunately, given the large number of model uncertainties, they are unable to make rigorous predictions for how long this phase may last.

We find the results of \citet{2017MNRAS.467.1154S} intriguing as they may shed light on another curious result of our study.
In the small galaxies ($r_d < 3\,\textrm{kpc}$) we observe a large scatter of the metallicity gradient, whilst larger galaxies present negative metallicity gradients with minimal scatter (Fig.~\ref{fig:grad_vs_size}).
If the smaller galaxies are comparatively less evolved systems, then this secular evolution (with metallicity gradient inversion at early times) may account for the diversity of metallicity gradients observed.
Consequently the larger galaxies, with their negative metallicity gradients, would be emblematic of the more classical understanding of inside-out growth which produces a common metallicity gradient \citep[e.g.][]{2000MNRAS.313..338P}.
This would reconcile our results with the low-redshift Universe.
And as such we could expect to find inverted metallicity gradients in small galaxies today.

The idea of that negative metallicity gradients only exist in galaxies with well-established discs is broadly similar to the findings of \citet{2017MNRAS.466.4780M}.
With their numerically simulated galaxies they find that galaxies with the steepest negative metallicity gradients all have well-ordered, rotating gas discs.
(N.B. by itself, this is a necessary, but insufficient condition.)
And conversely, the galaxies with perturbed discs all have flat metallicity gradients.

Nonetheless we strongly caution that their results are only conceptually similar to ours, not least because their work does not support a dependence between a galaxy's metallicity gradient and its size, which our work would suggest.

\section{Conclusions}\label{sec:conclusions}

Using MUSE data we present metallicity gradients for a sample of 84 intermediate-redshift galaxies ($0.1 \LA z \LA 0.8$).
We infer the true metallicity gradient using a forward-modelling technique that corrects for the seeing effects.
We search for trends of the observed metallicity gradient with global properties such as galaxy mass, size and SFR.
From this we conclude the following:

\begin{itemize}

\item The average galaxy in our sample has a negative metallicity gradient.
Nevertheless, there is significant scatter and we classify 7 of the 84 galaxies as having positive metallicity gradients.

\item We do not identify any significant correlation of the metallicity gradient with either total SFR or stellar mass (Figs.~\ref{fig:grad_vs_mass}~\&~\ref{fig:grad_vs_sfr}).
And we find no correlation whatsoever once the SFR is normalized relative to the main sequence (Fig.~\ref{fig:grad_vs_dsfr}).

\item This lack of correlation in the latter runs contrary to previous studies.
These studies, predominantly at higher redshifts ($0.6 \LA z \LA 2.6$), cite such a correlation as being suggestive of sudden gas inflow or merger events that might trigger star formation while simultaneously flattening/inverting the metallicity gradient.
Our results, however, do not support this interpretation at the intermediate redshifts we study.

\item The largest galaxies in our sample ($r_d > 3\,\textrm{kpc}$) are found to have almost exclusively negative metallicity gradients.
On the contrary, the smaller galaxies ($r_d < 3\,\textrm{kpc}$) present a range of metallicity gradients  (Fig.~\ref{fig:grad_vs_size}).
The small galaxies exhibit a larger intrinsic scatter in the metallicity gradient.

\item We liken the large galaxies as being similar to galaxies observed in the Universe today, where galaxies present a common metallicity gradient.
In addition, we speculate that the size dichotomy may be related to a secular understanding of inside-out growth, where a common metallicity gradient is only established in large (well-evolved) systems.

\item However, we advise caution on this last point.
Intrinsically smaller galaxies are also more affected by seeing-induced resolution loss.
Therefore, for these small galaxies our inferred metallicity gradients (and their errors) are presumably more dependent on the accuracy of our modelling (see Section~\ref{sec:sensitivity_analysis}).

\end{itemize}

The MUSE GTO surveys are currently ongoing and in the future will provide a larger sample of metallicity gradients.
This data will allow us to separate potential selection effects and biases.

In the future we will also study metallicity gradients in conjunction with the gas kinematics of the galaxies.
By also quantifying the local environments of our galaxies, we can study the impact of galaxy--galaxy interactions on the metallicity gradient, without relying on SFR as an indirect tracer.

\section*{Acknowledgements}

Thank you to Josephine Kerutt and Rikke Saust for their helpful discussions and providing PSF measurements for the CDFS fields.

This work was performed using the \textsc{python} and \textsc{idl} programming languages.
In particular we heavily used the following packages: \textsc{NumPy} \citep{doi:10.1109/MCSE.2011.37}, \textsc{SciPy} \citep{scipy}, \textsc{matplotlib} \citep{2007CSE.....9...90H} and \textsc{astropy} \citep{2013A&A...558A..33A}.

DC, JR, and VP acknowledge support from the ERC starting grant 336736-CALENDS.

JB acknowledges support by Funda\c{c}\~{a}o para a Ci\^{e}ncia e a Tecnologia (FCT) through national funds (UID/FIS/04434/2013) and by FEDER through COMPETE2020 (POCI-01-0145-FEDER-007672). During part of this work, JB was supported by FCT through Investigador FCT contract IF/01654/2014/CP1215/CT0003. 

TC acknowledges support of the ANR FOGHAR (ANR-13-BS05-0010-02), the OCEVU Labex (ANR-11-LABX-0060), and the A*MIDEX project (ANR-11-IDEX-0001-02) funded by the ``Investissements d'avenir'' French government program.

JS acknowledges support from the Netherlands Organisation for Scientific Research (NWO), through VICI grant 639.043.409, and the European Research Council under the European Union's Seventh Framework Programme (FP7/2007- 2013) / ERC Grant agreement 278594-GasAroundGalaxies. 




\bibliographystyle{mnras}
\bibliography{references.bib}



\appendix

\section{Additional figures and tables}

\begin{figure*}
\includegraphics[width=\linewidth]{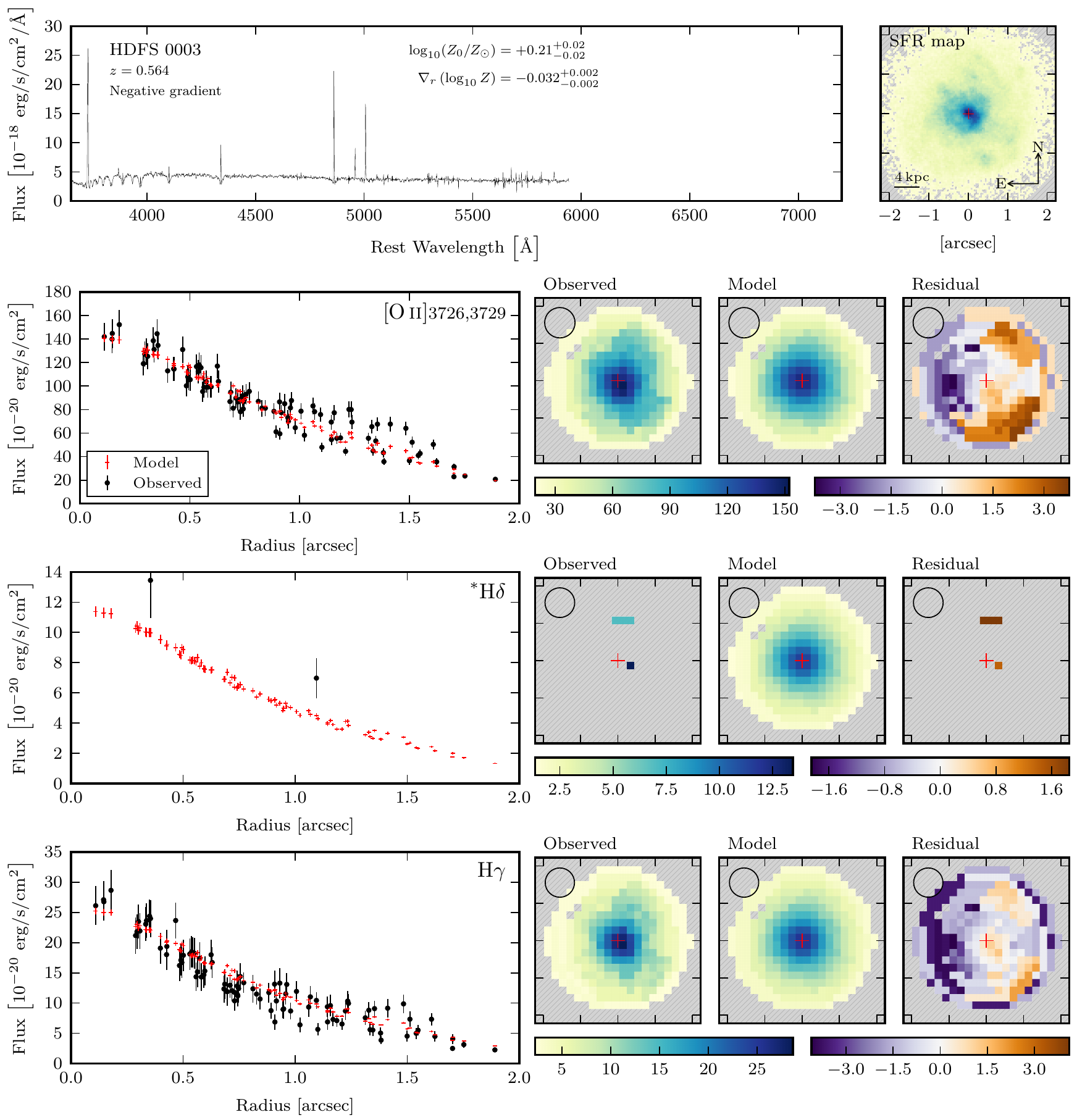}
\caption{Visual quality assessment of the model for galaxy HDFS-0003.
\emph{This is an illustrative example and similar plots for all galaxies can be found in the online supplementary material and.}
Top left: The globally integrated MUSE spectrum.
Also indicated is the derived central metallicity, $\log_{10}Z_0$ in dex, and the derived metallicity gradient, $\nabla_r\left(\log_{10}Z\right)$ in $\textrm{dex}/\textrm{kpc}$.
Top right: The SFR map that was used as input for the modelling.
A red cross indicate the galaxy centre.
Left: The radial flux profiles for the modelled emission lines.
The name of the emission line is indicated in the top right corner of each panel.
An asterisk denotes the emission lines that were not explicitly optimized by the spatial binning.
Consequently, these lines are often not observed at $\textrm{S/N} \ge 5$ in all spatial bins.
Black data points indicate observed fluxes and their $\pm1\sigma$ errors.
The red crosses show the median model solution.
The size of the vertical bar indicates the $\pm2\sigma$ range in fluxes.
Right: For each emission line we show three images.
These are, respectively, the 2D binned images of the observed fluxes, model fluxes, and scaled residuals $\left(\sfrac{\left(\rm \displaystyle Observed - Model\right)}{\displaystyle \rm Error}\right)$ for each emission line.
A black circle in the top left corner represents the FWHM of the PSF.
All images (including the SFR map) are shown on the same spatial scale.
[\emph{For space, this plot has been truncated to show only \{\forbidden{O}{ii}{}, \Hdelta{}, \Hgamma{}\}.
The full plot showing all lines (including also \{\Hbeta{}, \forbidden{O}{iii}{4959}, \forbidden{O}{iii}{5007}\}) is available in the supplementary material.}]
}
\label{fig:model_fit}
\end{figure*}

\begin{figure*}
\includegraphics[width=\linewidth]{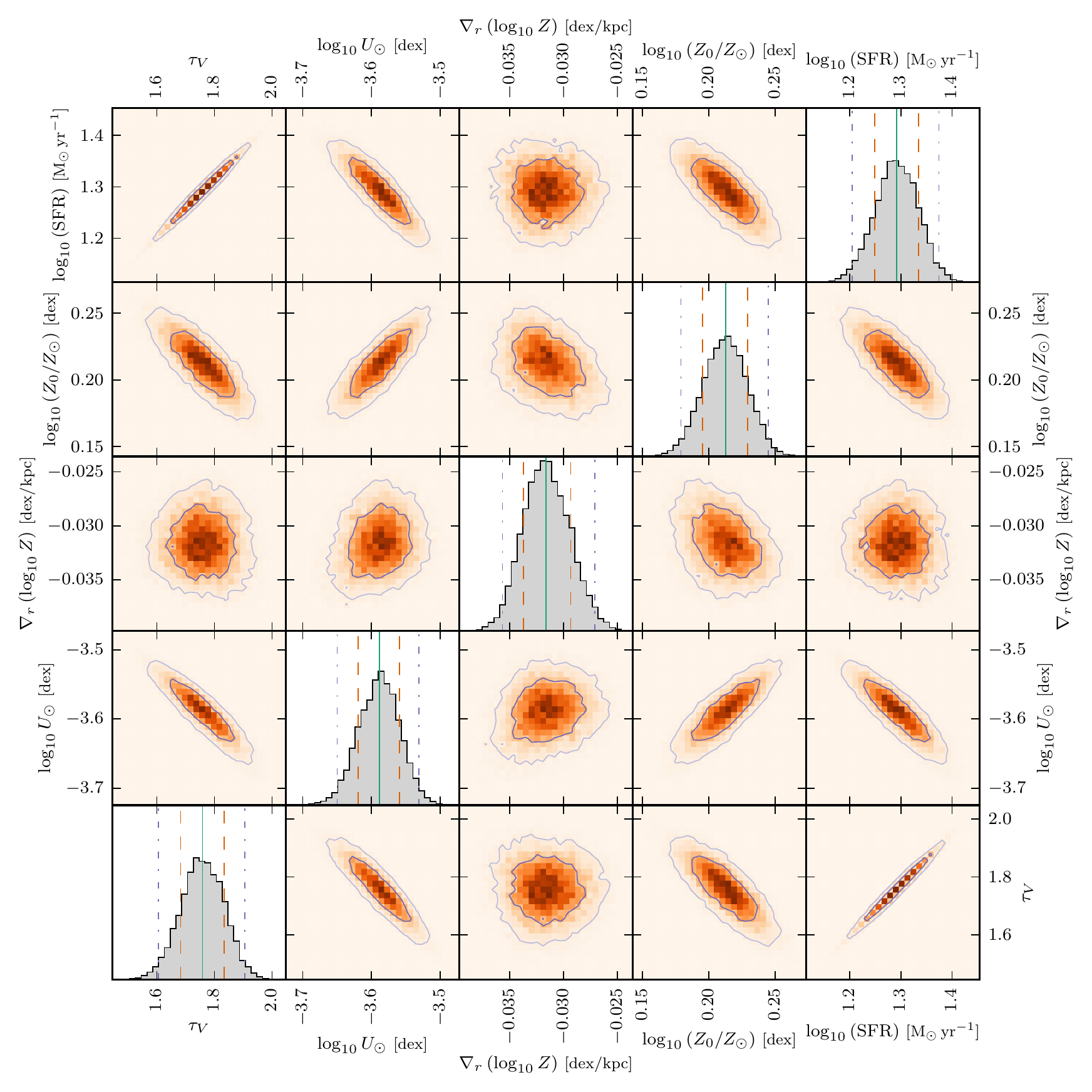}
\caption{Plot showing derived model parameters for HDFS 0003.
\emph{This is an illustrative example and similar plots for all galaxies can be found as online supplementary material.}
We show both 1D and 2D marginalized histograms for all 5 parameters: the total star-formation rate, $\log_{10}\left(\textrm{SFR}\right)$, central metallicity, $\log_{10}Z_0$, metallicity gradient, $\nabla_r\left(\log_{10}Z\right)$, ionization parameter at solar metallicity, $\log_{10}U_{\sun}$, and V-band optical depth, $\tau_V$.
In each 1D histogram the vertical lines indicate the median (solid), $\pm1\sigma$ quantiles (dashed) and $\pm2\sigma$ quantiles (dash-dotted).
In each 2D histogram we plot $1\sigma$ and $2\sigma$ contours.
All axes span a $[-4\sigma,4\sigma]$ interval in their respective parameters.
}
\label{fig:triangle}
\end{figure*}

In Fig.~\ref{fig:stamps} we provide an HST image atlas of our final galaxy sample.
With each cutout image we indicate the inferred metallicity gradient.
In Figs.~\ref{fig:model_fit}~\&~\ref{fig:triangle} we provide example plots for inspecting the quality of the model fit.
Similar plots for each galaxy in our final sample are provided in the online supplementary material.
In Table~\ref{tab:results} provide the derived metallicity gradients and other data pertaining to our galaxy sample.

\begin{figure*}
\includegraphics[width=\linewidth]{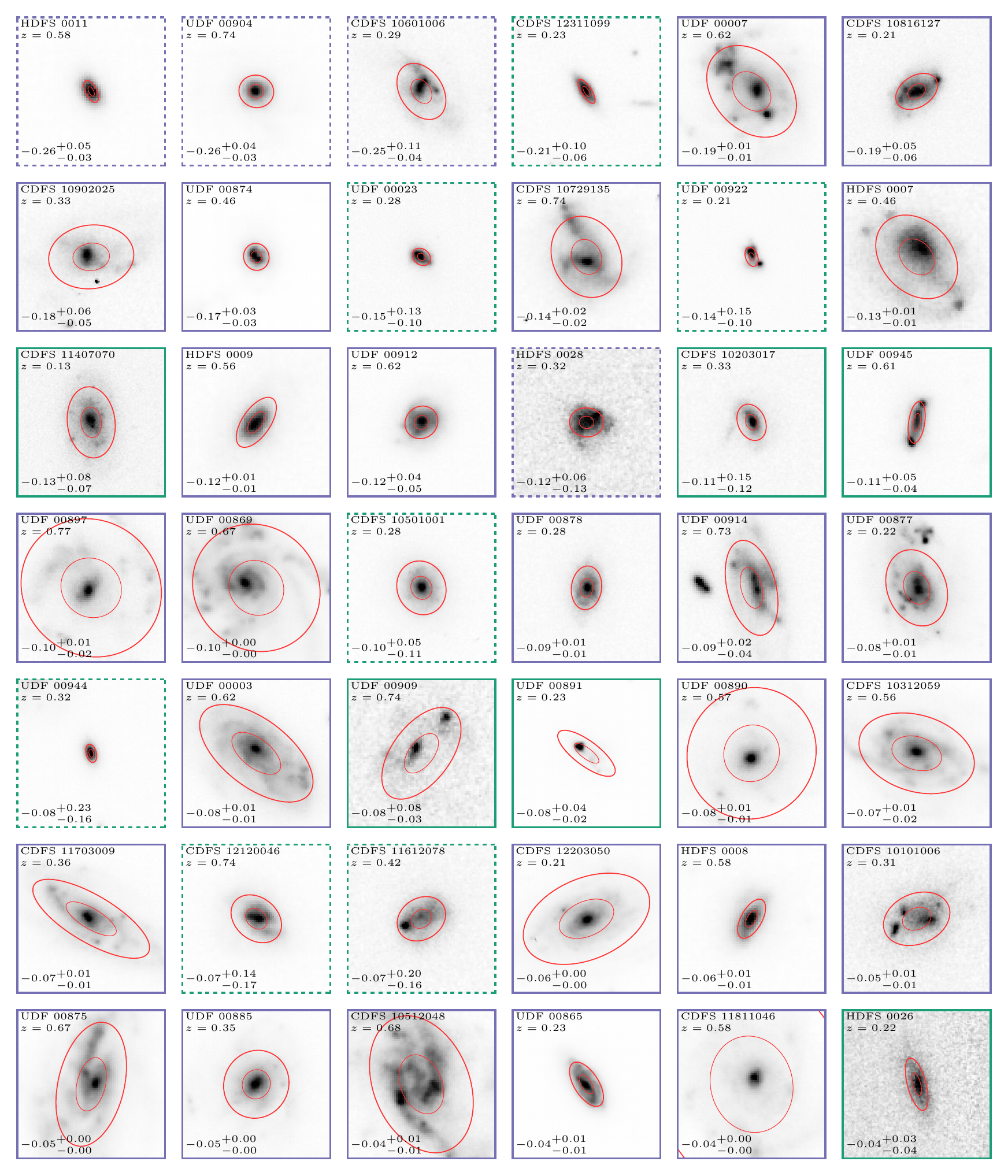}
\caption{Image atlas showing HST cutouts of galaxies in our sample, shown in order of increasing metallicity gradient.
For the UDF galaxies with display the F775W filter images, for all other galaxies we use the F814W filter.  
Each image is centred on the galaxy of interest, where North is up and East is left.
The image size is $16\,\textrm{kpc}$ (physical) edge-to-edge.
Two red ellipses indicate the derived galaxy morphology, drawn at the 50\% ($1.68\,r_d$) and 90\% ($3.89\,r_d$) light radii.
(Note that for the largest galaxies the 90\% radius lies outside of the cutout region shown.)
In the top-left corner of each panel we show the galaxy's redshift.
In the bottom-left corner we display the inferred metallicity gradient (units are $\textrm{dex}/\textrm{kpc}$).
The border colour displays the metallicity gradient type: negative gradients (blue), positive gradients (orange), flat gradients (green).
Galaxies flagged as unreliable have dashed borders.
}
\label{fig:stamps}
\end{figure*}

\begin{figure*}
\includegraphics[width=\linewidth]{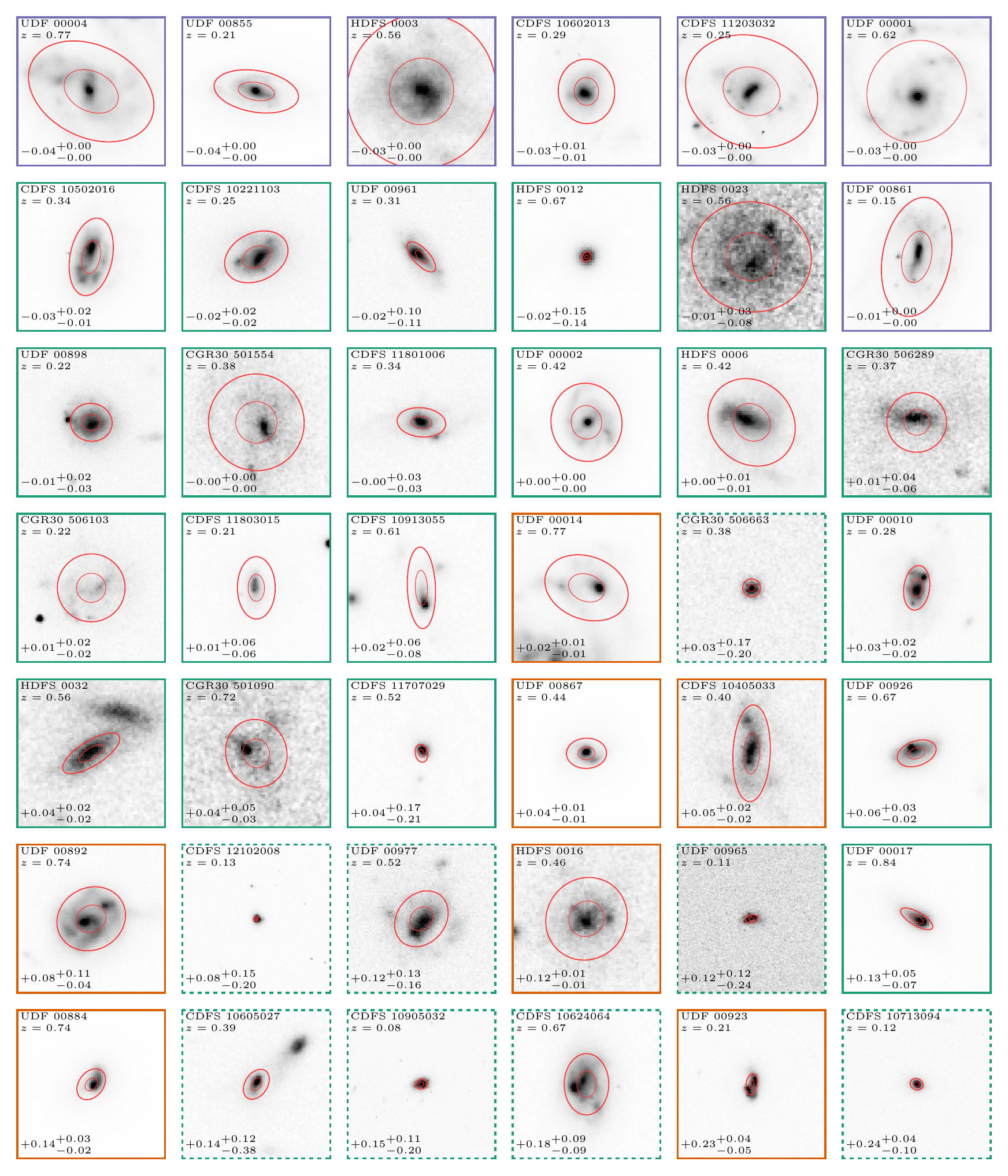}
\contcaption{}
\end{figure*}

\begin{landscape}
\begin{table}
\caption{Derived metallicity, mass, SFR and morphological properties.
Here we only show the first ten rows of the table.
The full table is available as part of the online supplementary material, and via the CDS archive (\url{http://cdsarc.u-strasbg.fr/viz-bin/qcat?J/MNRAS/478/4293}).
Columns are as follows.
Field: The targeted field (see Section~\ref{sec:field_desc}).
ID: Identification number of galaxy within the field.
$\log_{10}Z_0$: Median derived central metallicity.
$\nabla_r\left(\log_{10}Z\right)$: Median derived metallicity gradient.
Classification: Metallicity gradient classification (see Section~\ref{sec:results}).
$\log_{10}\left(\textrm{M}_\ast\right)$: Median stellar mass derived from \textsc{magphys}.
$\log_{10}\left(\textrm{SFR}\right)$: Median SFR derived from emission lines, uncertainties are from Monte-Carlo perturbation of fluxes.
$\Delta\textrm{SFR}$: Galaxy SFR normalized relative to a galaxy with identical mass on the main sequence.
RA: Right ascension of galaxy centre.
Dec.: Declination of galaxy centre.
$z$: Redshift derived from MUSE spectra.
$r_d$: Exponential disc scale length (see Section~\ref{sec:morphology}).
inc.: Galaxy inclination (see Section~\ref{sec:morphology}).
PA: Postition angle of galaxy's major axis on the sky, where $\textrm{North} = 0\degr$ and $\textrm{East} = 90\degr$.
All quoted uncertainties are the 16\textsup{th} and 84\textsup{th} percentiles.
}
\label{tab:results}

\def\arraystretch{1.4}
\begin{tabular}{cccccccccccccc}
\hline
Field & ID & $\log_{10}Z_0$ & $\nabla_r\left(\log_{10}Z\right)$ & Classification & $\log_{10}\left(\textrm{M}_\ast\right)$ & $\log_{10}\left(\textrm{SFR}\right)$ & $\Delta\textrm{SFR}$ & RA & Dec. & $z$ & $r_d$ & inc. & PA \\
 & &$[\textrm{dex}]$ & $[\textrm{dex}/\textrm{kpc}]$ & & $[\textrm{M}_{\sun{}}]$ & $[\textrm{M}_{\sun{}}\,\textrm{yr}^{-1}]$ & $[\textrm{dex}]$ & $[\textrm{deg}]$ & $[\textrm{deg}]$ & & $[\textrm{kpc}]$ & $[\textrm{deg}]$ & $[\textrm{deg}]$ \\
\hline
HDFS & 0003 & ${+0.21}_{-0.02}^{+0.02}$ & ${-0.032}_{-0.002}^{+0.002}$ & Negative & ${9.75}_{-0.07}^{+0.21}$ & ${+1.39}_{-0.27}^{+0.26}$ & $+0.90\pm 0.28$ & $338.223925$ & $-60.560437$ & $0.564$ & $4.3$ & $16.0$ & $-18.8$ \\
HDFS & 0006 & ${-0.20}_{-0.15}^{+0.16}$ & ${+0.002}_{-0.006}^{+0.005}$ & Flat & ${9.40}_{-0.13}^{+0.16}$ & ${-0.02}_{-0.29}^{+0.30}$ & $-0.13\pm 0.31$ & $338.242610$ & $-60.558759$ & $0.422$ & $2.6$ & $29.0$ & $+45.0$ \\
HDFS & 0007 & ${+0.68}_{-0.02}^{+0.01}$ & ${-0.133}_{-0.007}^{+0.007}$ & Negative & ${9.49}_{-0.19}^{+0.27}$ & ${+0.19}_{-0.33}^{+0.34}$ & $-0.03\pm 0.36$ & $338.247608$ & $-60.561020$ & $0.464$ & $2.6$ & $41.0$ & $+43.0$ \\
HDFS & 0008 & ${+0.37}_{-0.04}^{+0.04}$ & ${-0.058}_{-0.014}^{+0.013}$ & Negative & ${10.00}_{-0.10}^{+0.16}$ & ${+1.12}_{-0.28}^{+0.28}$ & $+0.46\pm 0.29$ & $338.214616$ & $-60.560443$ & $0.577$ & $1.2$ & $61.0$ & $-29.4$ \\
HDFS & 0009 & ${+0.45}_{-0.02}^{+0.02}$ & ${-0.124}_{-0.009}^{+0.009}$ & Negative & ${9.49}_{-0.15}^{+0.20}$ & ${+0.96}_{-0.28}^{+0.29}$ & $+0.63\pm 0.31$ & $338.233673$ & $-60.570605$ & $0.564$ & $1.6$ & $61.0$ & $-34.8$ \\
HDFS & 0011 & ${+0.38}_{-0.09}^{+0.07}$ & ${-0.261}_{-0.028}^{+0.052}$ & Unreliable & ${9.31}_{-0.12}^{+0.17}$ & ${+0.52}_{-0.29}^{+0.30}$ & $+0.29\pm 0.31$ & $338.218127$ & $-60.559180$ & $0.578$ & $0.7$ & $62.0$ & $+27.3$ \\
HDFS & 0012 & ${-0.24}_{-0.13}^{+0.16}$ & ${-0.023}_{-0.140}^{+0.148}$ & Flat & ${9.11}_{-0.12}^{+0.27}$ & ${+0.53}_{-0.36}^{+0.36}$ & $+0.32\pm 0.38$ & $338.237658$ & $-60.556392$ & $0.670$ & $0.3$ & $37.0$ & $+144.0$ \\
HDFS & 0016 & ${-0.38}_{-0.05}^{+0.04}$ & ${+0.119}_{-0.008}^{+0.010}$ & Positive & ${8.84}_{-0.13}^{+0.22}$ & ${+0.29}_{-0.26}^{+0.28}$ & $+0.48\pm 0.29$ & $338.230169$ & $-60.568745$ & $0.465$ & $2.4$ & $20.0$ & $-40.0$ \\
HDFS & 0023 & ${-0.14}_{-0.21}^{+0.29}$ & ${-0.013}_{-0.078}^{+0.032}$ & Flat & ${8.75}_{-0.17}^{+0.28}$ & ${-0.18}_{-0.49}^{+0.48}$ & $-0.05\pm 0.51$ & $338.232232$ & $-60.559392$ & $0.564$ & $3.3$ & $24.0$ & $+78.3$ \\
HDFS & 0026 & ${+0.03}_{-0.12}^{+0.12}$ & ${-0.037}_{-0.036}^{+0.031}$ & Flat & ${7.31}_{+0.00}^{+0.00}$ & ${-2.04}_{-0.03}^{+0.03}$ & $-0.52\pm 0.03$ & $338.232615$ & $-60.558401$ & $0.225$ & $1.5$ & $70.0$ & $+15.0$ \\
\hline
\end{tabular}

\end{table}
\end{landscape}


\bsp	
\label{lastpage}
\end{document}